Emmanuel Agullo, Camille Coti, Jack Dongarra, Thomas Herault, and Julien Langou.
\documentclass[conference,12pt,onecolumn]{IEEEtran}
\usepackage[dvips]{graphicx}
\graphicspath{{figures/}}
\usepackage[tight]{subfigure}
\usepackage{algorithm}
\usepackage{algorithmic}
\usepackage{url}
\usepackage{boxedminipage}
\usepackage{subfigure}

\usepackage{simplemargins}
\setpageheight{11in}
\setpagewidth{8.5in}

\setleftmargin{0.75in} 
\setrightmargin{0.75in}
\settopmargin{0.75in}
\setbottommargin{0.35in}

\newcommand{\ignore}[1]{{}}
\newtheorem{property}{Property}

\title{QR Factorization of Tall and Skinny Matrices in a
  Grid Computing Environment}

\author{ 
    \authorblockN{ 
          Emmanuel Agullo \authorrefmark{1}, 
          Camille Coti \authorrefmark{2}, 
          Jack Dongarra \authorrefmark{1},\\
         Thomas Herault \authorrefmark{3}, 
          Julien Langou \authorrefmark{4}}
  \authorblockA{\authorrefmark{1} Dpt of Electrical Engineering
    and Computer Science, University of Tennessee,\\ 
    1122 Volunteer Blvd, Claxton Building, Knoxville, TN 37996-3450, USA}
  \authorblockA{\authorrefmark{2}INRIA Saclay-\^Ile de France, F-91893
    Orsay, France} 
  \authorblockA{\authorrefmark{3}Univ Paris Sud, University of Tennesse, LRI,
    INRIA}
  \authorblockA{\authorrefmark{4} Dpt of Mathematical and
    Statistical Sciences, University of Colorado Denver,\\ 
    Campus Box 170, P.O. Box 173364, Denver, Colorado 80217-3364, USA}
  \authorblockA{eagullo@eecs.utk.edu, coti@lri.fr,
    dongarra@eecs.utk.edu, \\
    herault@lri.fr, julien.langou@ucdenver.edu }
}

\begin{document}

\maketitle

\footnotetext[1]{This work was partly supported by the EC grant for the
  QosCosGrid project (grant \# FP6-2005-IST-5 033883), and NSF-CCF (grant \#881520).}

\begin{abstract}
Previous studies have reported that common dense linear algebra
operations do not achieve speed up by using multiple geographical sites
of a computational grid. Because such operations are the building
blocks of most scientific applications, conventional supercomputers
are still strongly predominant in high-performance computing and the
use of grids for speeding up large-scale scientific problems is
limited to applications exhibiting parallelism at a higher level. We
have identified two performance bottlenecks in the distributed memory
algorithms implemented in ScaLAPACK, a state-of-the-art dense linear
algebra library. First, because ScaLAPACK assumes a homogeneous
communication network, the implementations of ScaLAPACK algorithms
lack locality in their communication pattern. Second, the number of
messages sent in the ScaLAPACK algorithms is significantly greater
than other algorithms that trade flops for
communication. In this paper, we present a new approach for computing
a QR factorization -- one of the main dense linear algebra kernels --
of tall and skinny matrices in a grid computing environment that
overcomes these two bottlenecks. Our contribution is to articulate a
recently proposed algorithm (Communication Avoiding QR) with a
topology-aware middleware (QCG-OMPI) in order to confine intensive
communications (ScaLAPACK calls) within the different geographical
sites. An experimental study conducted on the Grid'5000 platform shows
that the resulting performance increases linearly with the number of
geographical sites on large-scale problems (and is in particular
consistently higher than ScaLAPACK's).
\end{abstract}

\section{Introduction}\label{sec:intro}

Grid computing~\cite{foster_grid_blueprint} as a utility has reached
the mainstream. Many large-scale scientific problems have been
successfully solved thanks to the use of computational grids (or,
simply, \emph{grids}). These problems cover a wide range of scientific
disciplines including biology (protein
folding~\cite{Larson_folding@homeand}), medicine (cure muscular
dystrophy~\cite{raphael_bolze_thesis}), financial modeling, earthquake
simulation, and climate/weather modeling. Such scientific
breakthroughs have relied on the tremendous processing power provided
by grid infrastructures. For example, the Berkeley Open
Infrastructure for Network Computing (BOINC)~\cite{BOINC} gathers
the processing power of personal computers provided by people
volunteering all over the world. This processing power is then made
available to researchers through different projects such as
Climateprediction.net~\cite{climateprediction},
Rosetta@home~\cite{rosetta} and World Community Grid
(WCG)\footnote{http://www.worldcommunitygrid.org}. As of September 2009, 18, BOINC had 566,000 active
computers (hosts) worldwide for an average total processing power of
$2.4$~Pflop/s\footnote{\url{http://boincstats.com/}}.
Furthermore, following the supercomputing trends, grid computing
infrastructures have successfully exploited the emerging hardware technologies.
The Folding@Home project~\cite{folding@home} -- which aims at understanding
protein folding, misfolding, and related diseases -- achieves $7.9$~Pflop/s
thanks to grid exploiting specialized hardware such as graphics processing
units (GPUs), multicore chips and IBM Cell processors.


However, conventional supercomputers are strongly predominant in
high-performance computing (HPC) because different limiting factors
prevent the use of grids for solving large-scale scientific problems.
First of all, security requirements for grids are not completely met
in spite of the important efforts in that
direction~\cite{ieee_security_for_grids}. Second,
contrary to their original purpose (the term \emph{grid} itself is a
metaphor for making computer power as easy to access as an electric
power grid~\cite{foster_grid_blueprint}), grids have not been historically
very user-friendly. Third, not all the grid infrastructures are
dimensioned for HPC, which is only one of the aims of grid
computing. Even recent commercial offerings such as Amazon Elastic
Compute Cloud (EC2)\footnote{http://aws.amazon.com/ec2/} are not considered mature yet
for HPC because of under-calibrated
components~\cite{Walker08}. Furthermore, other aspects are still the
focus of intensive research, such as service discovery~\cite{p2p_sd},
scheduling~\cite{GridWay08}, \emph{etc.}

But, above all, the major limiting factor to a wider usage of grids by
computational scientists to solve large-scale problems is the fact
that common dense linear algebra operations do not achieve performance
speed up by using multiple geographical sites of a computational grid,
as reported in previous studies~\cite{gridtop500,grads}. Because
those operations are the building blocks of most scientific
applications, the immense processing power delivered by grids
vanishes. Unless the application presents parallelism at a higher
level (most of the applications running on BOINC are actually
\emph{embarrassingly parallel}, \emph{i.e.}, loosely coupled), its
performance becomes limited by the processing power of a single
geographical site of the grid infrastructure, ruining the ambition to
compete against conventional supercomputers. We have identified two
performance bottlenecks in the distributed memory algorithms
implemented in ScaLAPACK~\cite{scalapack}, a state-of-the-art dense
linear algebra library. First, because ScaLAPACK assumes a homogeneous
communication network, the implementations of the ScaLAPACK algorithms
lack locality in their communication pattern. Second, the number of
messages sent in the ScaLAPACK algorithms is significantly greater
than other algorithms that trade flops for
communication. In this paper, we present a new approach for
factorizing a dense matrix -- one of the most important operations
in dense linear algebra -- in a grid computing environment that
overcomes these two bottlenecks. Our approach consists of articulating
a recently proposed algorithm (Communication Avoiding
algorithm~\cite{CAQR}) with a topology-aware middleware
(QCG-OMPI\cite{CHPRC08}) in order to confine intensive
communications (ScaLAPACK calls) within the different geographical sites.

In this study, we focus on the QR factorization~\cite{Golub89} of a
tall and skinny (TS) dense matrix into an orthogonal matrix $Q$ and an
upper triangular matrix $R$ and we discuss how our approach
generalizes to all one-sided factorizations (QR, LU and Cholesky) of a
general dense matrix (Section~\ref{sec:model}). Furthermore, we focus
on the computation of the triangular factor $R$ and do not explicitly
form the orthogonal matrix $Q$. However, we show that the performance
behavior would be similar if we compute $Q$ or not.

The paper is organized as follows. We present the related work and
define the scope of our paper in Section~\ref{sec:relatedwork}. In
Section~\ref{sec:contribution}, we present the implementation of a QR
factorization of TS matrices that confines intensive communications
within the different geographical sites. Section~\ref{sec:model}
discusses a performance model that allows us to understand the basic
trends observed in our experimental study (Section~\ref{sec:xp}). We
conclude and present the future work in Section~\ref{sec:conclusion}.

\ignore{

Based on orthogonal transformations, this method is numerically stable
and is a first step toward the resolution of least square
systems~\cite{Golub89}.

An experimental study conducted on the Grid'5000 platform shows that
the resulting performance increases linearly with the number of
geographical sites on large-scale problems (and is in particular
consistently higher than ScaLAPACK's).

 To address the question
``Can cloud computing reach the TOP500 ?'', \cite{gridtop500} studied
the performance of the factorization of a dense matrix.

HeteroScalapack, Grads.

*** Plan: related work (other dense linear algebra applications),
related work, our approach, experimental envirnoment,
tuning, experiments.

-------------------------

loosely coupled
slow inerconnect

QR factorization is one of the major one-sided factorizations in dense
linear algebra. Based on orthogonal transformations, this method is
well known to be numerically stable and is a first step toward the
resolution of least square systems~\cite{Golub89}.

The purpose of this paper is to. Communication-Avoiding QR
(CAQR)~\cite{demmel-tsqr} algorithm introduced by Demmel et al. CAQR
factors general rectangular distributed matrices with a parallel panel
factorization.

}

\section{Background}\label{sec:relatedwork}

We present here the related work. We first describe previous
experimental studies of the behavior of dense linear algebra
operations in a grid computing environment
(Section~\ref{sec:lagrid}). We then succinctly present the operation
we focus on in this paper, the QR factorization, as it is implemented
in ScaLAPACK, a state-of-the-art dense linear algebra library for
distributed memory machines (Section~\ref{sec:qrfacto}). We continue
with the introduction of a recently proposed algorithm trading flops
for communication (Section~\ref{sec:caqr}). To take advantage in a
grid computing environment of the limited amount of communication
induced by such an algorithm, we need to articulate it with
the topology of the grid. We present in Section~\ref{sec:qcg-ompi} a
middleware enabling this articulation by (i) retrieving the system
topology to the application and even (ii) allowing the application to
reserve suitable resources. Such an articulation of the algorithms
with the topology is critical in an environment built on top of
heterogeneous networks such as a grid. We conclude this review by
discussing the scope of this paper (Section~\ref{sec:scope}).


\subsection{Dense linear algebra on the grid}
\label{sec:lagrid}


The idea of performing dense linear algebra operations on the grid is
not new; however, success stories are rare in the related bibliography.
Libraries that have an MPI~\cite{MPI} interface for handling the communication layer,
such as ScaLAPACK or HP Linpack,
can be run on a grid by
linking them to a grid-enabled implementation of the MPI standard such
as MPICH-G2~\cite{MPICHG2}, PACX-MPI~\cite{PACX} or GridMPI\footnote{http://www.gridmpi.org}.
MPI has
become the \textit{de facto} language for programming parallel
applications on distributed memory architectures such as
clusters.
Programmers have gained experience using this programming
paradigm throughout the past decade; scientific libraries have been
developed and optimized using MPI. As a consequence, it is natural to
consider it as a first-choice candidate for programming parallel
applications on the grid in order to benefit from this experience and
to be able to port existing applications for the grid.
 The 
GrADS\footnote{Software Support for High-Level Grid Application
  Development \url{http://www.hipersoft.rice.edu/grads/}} project had
the purpose of simplifying distributed, heterogeneous computing and
making grid application development as well as performance tuning for
real applications an everyday practice. Among other accomplishments, 
large matrices could be factorized thanks to the use of a grid whereas
it was impossible to process them on a single cluster because of memory
constraints~\cite{grads, grads-self}. The resource allocation (number
of
clusters, \emph{etc.}) was automatically chosen in order to maximize the
performance. However, for matrices that could fit in the (distributed)
memory of the nodes of a cluster, the experiments (conducted with
ScaLAPACK) showed that the use of a single cluster was
optimal~\cite{grads}. In other words, using multiple geographical sites
led to a slow down of the factorization.  Indeed, the overhead due
to the high cost of inter-cluster communications was not balanced by
the benefits of a higher processing power.

For the same reason, the EC2 cloud has recently been shown to be inadequate
for dense linear algebra~\cite{gridtop500}. In this latter study, the
authors address the question whether cloud computing can reach the
Top500\footnote{http://www.top500.org}, \emph{i.e.}, the ranked list of the fastest
computing systems in the world. Based on experiments conducted
with the parallel LU factorization~\cite{Golub89} implemented in the
HP Linpack Benchmark~\cite{Linpack-benchmark}, not only did they observe a
slow down when using multiple clusters, but they also showed that the
financial cost (in dollars) of performance (number of floating-point
operations per second, in Gflop/s) increases exponentially with the
number of computing cores used, much in contrast to existing scalable
HPC systems such as supercomputers.

The HeteroScaLAPACK
project\footnote{\url{http://hcl.ucd.ie/project/HeteroScaLAPACK}} aims at
developing a parallel dense linear algebra package for heterogeneous
architectures on top of ScaLAPACK. This approach is orthogonal (and
complementary) to ours since it focuses on the heterogeneity of the
processors~\cite{HeteroSCALAPACK}, whereas we presently aim at mapping the
implementation of the algorithm to the heterogeneity of the network (topology)
through QCG-OMPI. In our present work, we do not consider the heterogeneity of the processors.
Another fundamental difference with HeteroScaLAPACK is that we are using TSQR, an algorithm that is not available in ScaLAPACK.

\subsection{ScaLAPACK's QR factorization}
\label{sec:qrfacto}

The QR factorization of an $M\times N$ real matrix $A$ has the form $A = QR$,
where $Q$ is an $M\times M$ real orthogonal matrix and R is an $M\times N$ real
upper triangular matrix. Provided $A$ is nonsingular, this factorization is
essentially unique, that is, it is unique if we impose the diagonal entries of
$R$ to be positive.  There is a variety of algorithms to obtain a QR
factorization from a given matrix, the most well-known arguably being the Gram-Schmidt algorithm.
Dense linear algebra libraries have been traditionally focusing on algorithms
based on unitary transformations (Givens rotations or Householder reflections)
because they are unconditionally backward stable~\cite{Golub89}. Givens rotations are advantageous
when zeroing out a few elements of a matrix whereas Householder transformations are advantageous
when zeroing out a vector of a matrix. Therefore, for dense matrices, we consider the QR
factorization algorithm based on Householder reflections.
The algorithm consists 
of applying successive elementary Householder transformations of the
form $H=I-\tau v v^T$ where $I$ is the identity matrix, $v$ is a
column reflector and $\tau$ is a scaling factor~\cite{Golub89}.

To achieve high performance on modern computers with different levels of cache,
the application of the Householder reflections is \emph{blocked}~\cite{schreiber1989storage}.
In ScaLAPACK~\cite{ScaLAPACK_1997_guide}, $b$ elementary Householder matrices are
accumulated within a \emph{panel} (a block-column) $V$ consisting of $b$ reflectors.
The consecutive applications of these $b$ reflectors ($H_1H_2...H_{b}$) is then constructed all at once using the matrix equality
$H_1H_2...H_{b}=I-VTV^T$
($T$ is a $b\times b$ upper triangular matrix).
However, this blocking incurs an additional computational overhead. The overhead is
negligible when there is a large number of columns
to be updated but is significant when there are only a few columns to be
updated. Default values in the ScaLAPACK PDGEQRF subroutine are NB=64 and NX=128,
where NB is the block size, $b$, and
NX is the cross-over point; blocking is not to be used if there is less than NX columns are to be updated.
PDGEQRF uses PDGEQR2 to perform the panel factorizations.
Due to the panel factorization,
the algorithm in ScaLAPACK requires one allreduce operation for each column of
the initial matrix. In other words, ScaLAPACK uses at least $N\log_2(P)$ messages
to factor an $M$-by--$N$ matrix.



\subsection{Communication Avoiding QR (CAQR) factorization}
\label{sec:caqr}

In this paper we propose an implementation of the so-called ``Communication
Avoiding QR'' (CAQR) algorithm originally proposed by Demmel et al.~\cite{CAQR}. CAQR belongs to the class of the
\emph{(factor panel) / (update trailing matrix)} algorithms.  For all algorithms in this class,
the update phase is entirely dictated by the panel factorization
step and is easily parallelizable. Therefore, we only discuss the panel factorization step.
The panel factorization in CAQR is based on
the ``Tall and Skinny QR'' factorization algorithm (TSQR)~\cite{CAQR}.  In contrast to the
ScaLAPACK panel factorization algorithm (subroutine PDGEQR2), which requires one
allreduce operation per column, TSQR requires one allreduce operation per $b$
columns where $b$ is an arbitrary block size. The number of communications is
therefore divided by $b$. The volume of communication stays the same. The
number of operations on the critical path is increased in TSQR by an additional
$\mathcal{O}(\log_2(P)N^3)$ term.  TSQR effectively trades communication for flops.

As explained in~\cite{CAQR}, TSQR is a single complex allreduce
operation. The TS matrix is split in $P$ block-rows, called
\emph{domains}; the factorization of a domain is the operation
performed on the leaves of the binary tree associated to the reduction.
The basic operation then used in this allreduce operation
is as follows: from two input triangular matrices $R_1$ and $R_2$,
stack $R_1$ on top of $R_2$ to form $[R_1;R_2]$, perform the QR
factorization of $[R_1;R_2]$, the output $R$ is given by the R-factor
of $[R_1;R_2]$.  One can show that this operation is binary and
associative. It is also commutative if one imposes the diagonal of
each computed R-factor to have nonnegative entries.  As for any reduce
operation, the shape of the optimal tree depends on the dimension of
the data and the underlying hardware. CAQR with a binary tree has been
studied in the parallel distributed context~\cite{CAQR} and CAQR with
a flat tree has been implemented in the context of out-of-core QR
factorization~\cite{gunter2005parallel}.  We note that CAQR with a
flat tree also delivers wide parallelism and, for this reason, has
been used in the multicore
context~\cite{buttari2007class,kurzak2008qr,quintana-orti2008scheduling}.

\begin{figure}[htb] 
\begin{minipage}[c]{.45\linewidth}    
        \begin{center}
            \includegraphics[width=\textwidth]{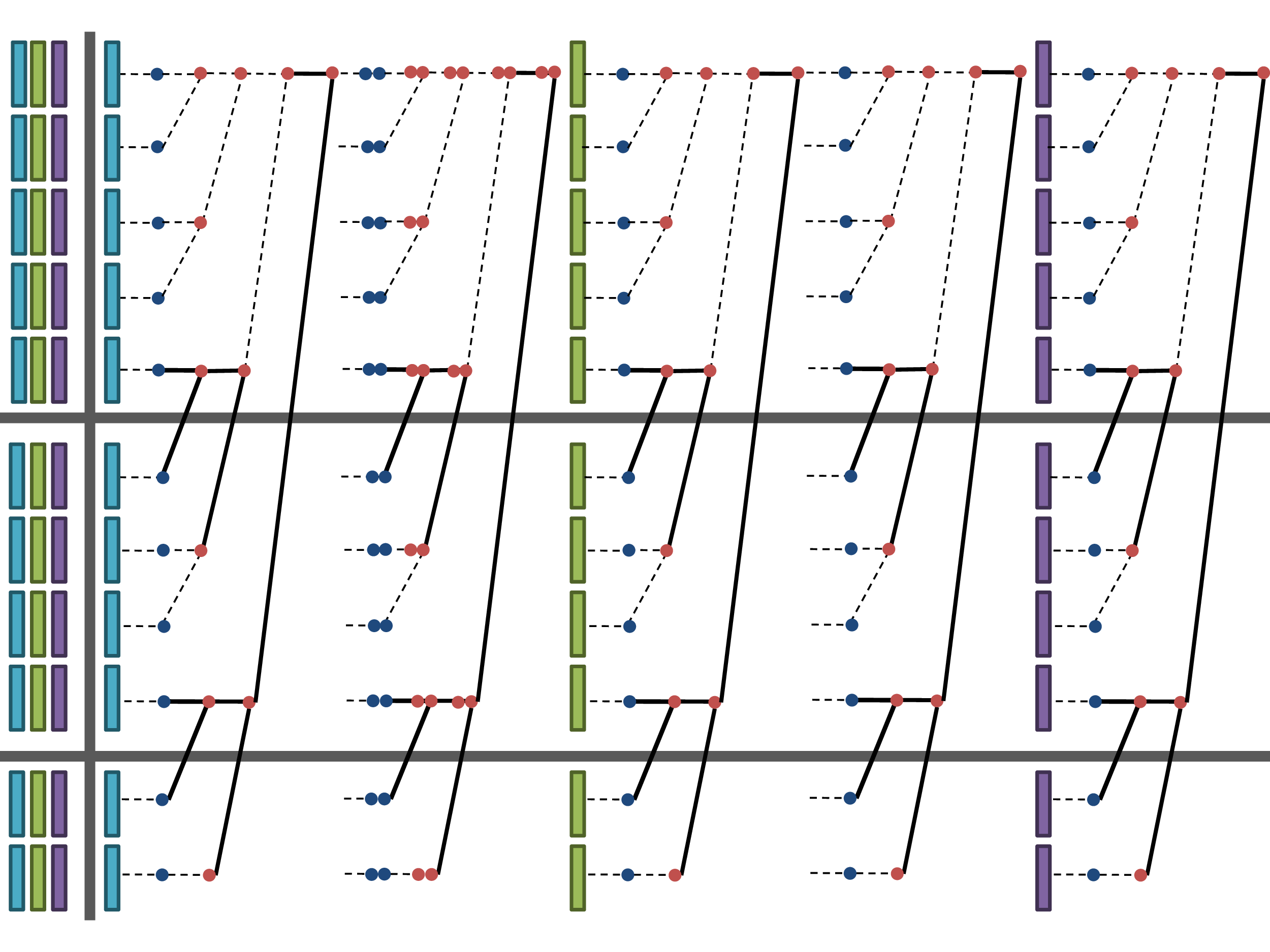}
        \end{center}
         \vspace{-0.5cm}
          \caption{
        \label{fig:figure_de_julien_qr2}
 \footnotesize{
Illustration of the
ScaLAPACK panel factorization routine on a $M$-by-3 matrix. It involves one reduction
per column for the normalization and one reduction per column for the update. (No update for the last column.)
The reduction tree used by ScaLAPACK is a binary tree. It
In this example, 
we have 25 inter-cluster messages (10 for all columns but the
last, 5 for the last). 
A tuned reduction tree would have given 10 inter-cluster messages (4 per column but the last, 2 for the last).
We note that if process ranks are randomly distributed, the figure can be worse.
}}
\end{minipage}\hfill
    \begin{minipage}[c]{.45\linewidth}
    \begin{center}
            \includegraphics[width=\textwidth]{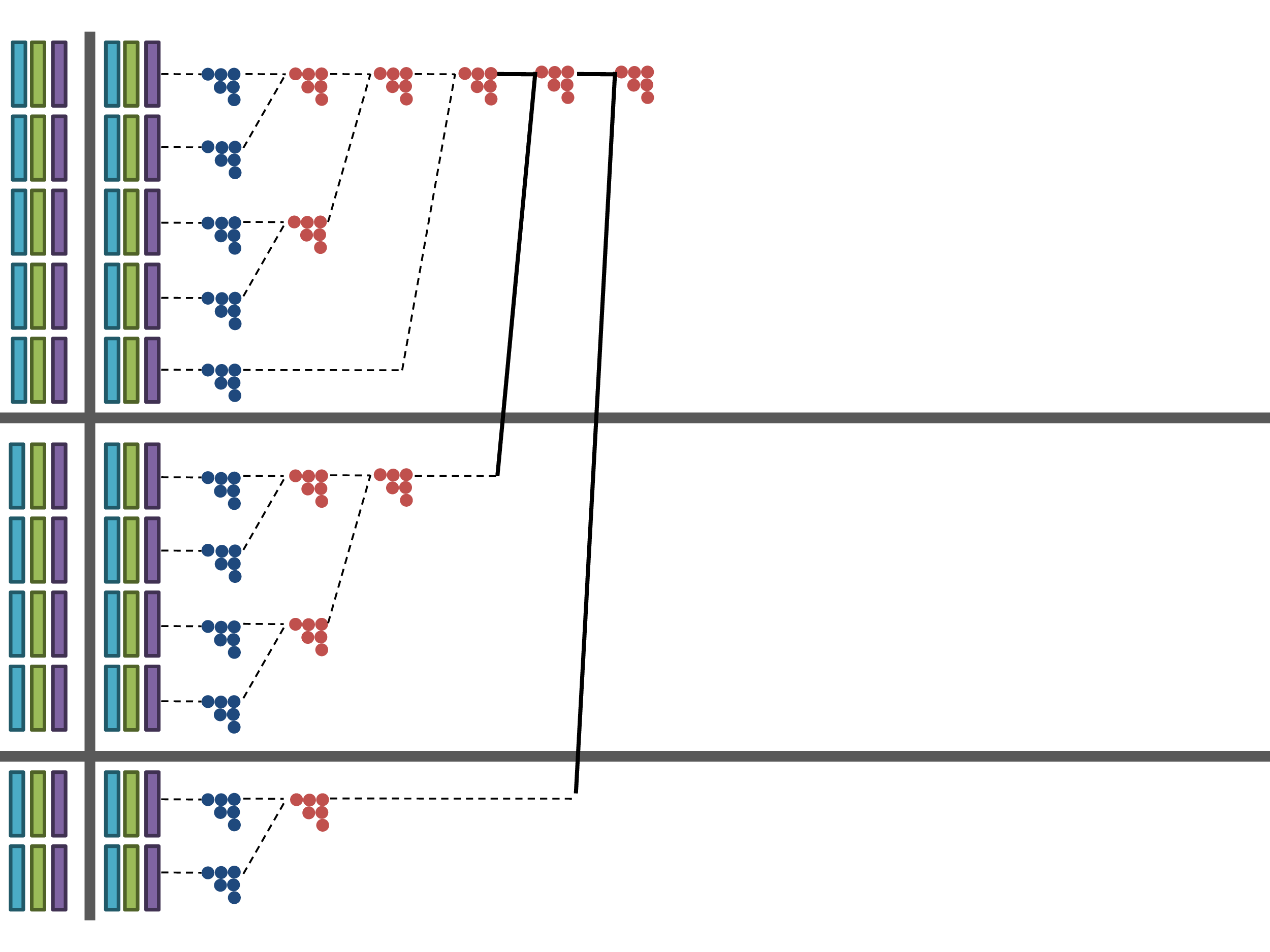}
        \end{center}        
        \vspace{-0.5cm}
        \caption{
        \label{fig:figure_de_julien_tsqr}
 \footnotesize{
Illustration of the
TSQR panel factorization routine on a $M$-by-3 matrix. It involves only one reduction tree.
Moreover the reduction tree 
is tuned for the grid architecture. We only have two inter-cluster messages. This 
number (two) is independent of the number of columns. This number is obviously optimal. One can not expect less than two
inter-cluster communications when data is spread on the three clusters.
~~~~}}
    \end{minipage}\hfill
\end{figure}

Previous implementations of CAQR have used either a flat tree or a binary tree.
One key originality of our present work lies in the fact that our reduction tree is neither binary nor flat.
It 
is tuned for the targeted computational grid, as illustrated in
Fig.~\ref{fig:figure_de_julien_tsqr}. First we reduce with a binary tree on each
cluster.  Then we reduce with a second binary tree the result of each cluster
at the grid level. The binary tree used by ScaLAPACK PDGEQR2
(Fig.~\ref{fig:figure_de_julien_qr2}) minimizes the sum of the inter-cluster messages and the intra-cluster messages. Our tree
is designed to minimize the
total number of inter-cluster messages.

We now give a brief history of related algorithmic work in contrast to the reference
work of Demmel et al.~\cite{CAQR}.  The parallelization of the Givens rotations
based and Householder reflections based QR factorization algorithms is a
well-studied area in Numerical Linear Algebra.  The development of the
algorithms has followed architectural trends.  In the late 1970s / early
1980s~\cite{heller:1978,LordKowalikKumar:1983,SamehDuck:1978}, the research
was focusing on algorithms based on Givens rotations. The focus was on 
extracting as much parallelism as possible. We can interpret these sequences of
algorithms as scalar implementations using a flat tree of the algorithm in Demmel et
al.~\cite{CAQR}.  In the late 1980s, the research shifted
gears and presented algorithms based on Householder
reflections~\cite{Pothen,ChuGeorge:1990}. The motivation was to use vector
computer capabilities.  We can interpret all these algorithms as vector
implementations using a flat tree and/or a binary tree of the algorithm in
Demmel et al.~\cite{CAQR}. All these algorithms require a number of messages
greater than $n$, the number of columns of the initial matrix $A$, as in ScaLAPACK.  The
algorithm in Demmel et al.~\cite{CAQR} is a generalization with multiple blocks
of columns with a nontrivial reduction operation, which enables one to divide the number of messages of these previous
algorithms by the block size, $b$.
Demmel et al. proved that TSQR and CAQR algorithms induce a minimum amount of
communication (under certain conditions, see Section 17 of~\cite{CAQR} for more
details) and are numerically as stable as the Householder QR factorization.


\subsection{Topology-aware MPI middleware for the grid: QCG-OMPI}
\label{sec:qcg-ompi}



\begin{figure}[htb]
  \subfigure[Communications performance on Grid'5000]{
    \begin{minipage}{0.55\linewidth}
      \small
      \begin{tabular}[b]{|l|l|l|l|l|l|}
        \hline
        Latency (ms) & Orsay & Toulouse & Bordeaux & Sophia \\ \hline
        Orsay & 0.07 & 7.97 & 6.98 & 6.12 \\ \hline
        Toulouse & ~ & 0.03 & 9.03 & 8.18 \\ \hline
        Bordeaux & ~ & ~ & 0.05 & 7.18 \\ \hline
        Sophia & ~ & ~ & ~ & 0.06 \\ \hline
        \hline
        Throughput (Mb/s) & Orsay & Toulouse & Bordeaux & Sophia \\ \hline
        Orsay & 890 & 78 & 90 & 102 \\ \hline
        Toulouse & ~ & 890 & 77 & 90 \\ \hline
        Bordeaux & ~ & ~ & 890 & 83 \\ \hline
        Sophia & ~ & ~ & ~ & 890 \\ \hline
      \end{tabular}
      \vspace{1.3em}
      \label{tab:comm}
      \end{minipage}
    }
 \subfigure[Grid'5000: a nation-wide experimental testbed.]{
    \begin{minipage}{0.35\linewidth}
     \includegraphics[width=0.85\linewidth]{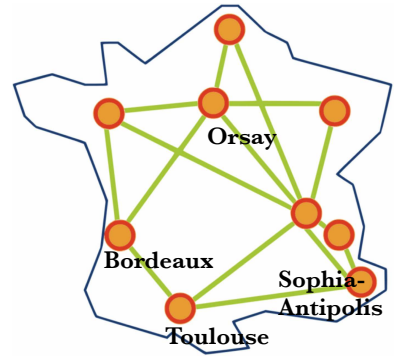}
   \end{minipage}
  }
  \caption{Grid'5000 communication characteristics.}
\end{figure}

Programming efficient applications for grids built by federating
clusters is challenging, mostly because of the difference of
performance between the various networks the application has to
use. As seen in the table of Figure~\ref{tab:comm} we can observe two orders of magnitude between inter and intra-cluster latency on a dedicated, nation-wide network, and the difference can reach three or four orders of magnitude on an international, shared network such as the Internet. As a consequence, the application must be adapted to the intrinsically hierarchical topology of the grid. In other words, the communication and computation patterns of the application must match the physical topology of the hardware resources it is executed on.
 
ScaLAPACK, and many of the linear algebra libraries for scientific
computing, are programmed in MPI.
MPI is fit for homogeneous supercomputers: processes 
are mostly indistinguishable one from another, and the standard does 
not specify anything about process / node placement. 

As a consequence, to efficiently program a parallel application on top
of a non-uniform network, typically on top of a hierarchical network
like a grid, MPI must be extended to help programmers adapt the
communications of the application to the
machine. MPICH-G2~\cite{MPICHG2} introduced the concept of colors to
describe the available topology to the application at runtime. Colors
can be used directly by MPI routines in order to build topology-aware
communicators (the abstraction in MPI that is used to group processors
together). However, the application is fully responsible to adapt
itself to the topology that is discovered at runtime. This adaptation,
and the load-balancing that it implies, may be a hard task for the
application.

The QosCosGrid system\footnote{Quasi-Opportunistic Supercomputing for Complex Systems in Grid Environments, http://www.qoscosgrid.eu} offers a resource-aware grid meta-scheduler that gives the possibility to allocate resources that match requirements expressed in a companion file called the application's \textit{JobProfile} that describe the future communications of the application \cite{BCGHKSS09}. The \emph{JobProfile} defines process groups and requirements on the hardware specifications of the resources that have to be allocated for these processes such as amount of memory, CPU speed, and network properties between groups of processes, such as latency and bandwidth.

As a consequence, the application will always be executed on an appropriate resource topology. It can therefore be developed for a specific topology in mind, for example, under the assumption that a set of processes will be located on the same cluster or on the same multi-core machine. Of course, the more flexibility the programmer gives to the JobProfile and the application, the more chances he gets to let the meta-scheduler find a suitable hardware setup.

The QosCosGrid system features QCG-OMPI, an MPI implementation based
on OpenMPI~\cite{openmpi} and targeted to computational grids. Besides
grid-specific communication features that enable communicating
throughout the grid described in~\cite{CHPRC08}, QCG-OMPI has the
possibility to retrieve topology information provided to the scheduler
in the JobProfile at run-time.
We explain in Section~\ref{sec:contribution} how we have implemented
and articulated TSQR with QCG-OMPI in order to take
advantage of the topology.

\subsection{Scope}
\label{sec:scope}


The QR factorization of TS matrices is directly used as a kernel
in several important applications of linear algebra.  For instance,
block-iterative methods need to regularly perform this operation in order to obtain
an orthogonal basis for a set of vectors; this step is of particular
importance for block eigensolvers (BLOPEX, SLEPc, PRIMME).
Currently these packages rely on unstable orthogonalization schemes to avoid
too many communications. TSQR is a stable algorithm that enables the same total
number of messages.
TSQR can also be used to perform the panel factorization of an algorithm
handling general matrices (CAQR). Thanks to simulations, Demmel et
al.~\cite{CAQR} anticipated that the benefits obtained with TSQR should get
transposed to CAQR. Said differently, this present study can be viewed as a
first step towards the factorization of general matrices on the grid.

Grids aggregate computing power from any kind of resource.  However, in 
some typical grid projects, such as Superlink@Technion, the Lattice project, 
EdGES, and the Condor pool at Univ. of Wisconsin-Madison, a 
significant part of the power comes from a few institutions featuring 
environments with a cluster-like setup. In this first work, we focus our 
study on clusters of clusters, to enable evaluation in a stable and 
reproducible environment. Porting the work to a general desktop 
grid remains a future work.

Finally, we emphasize that the objective of this paper is to show that
we can achieve a performance speed up over the grid with common dense
linear algebra operations. To illustrate our claim, we 
compare our approach against a state-of-the-art library for distributed memory
architectures, ScaLAPACK. In order to highlight the differences, we
chose to base our approach on ScaLAPACK (see
Section~\ref{sec:contribution}).

\section{QCG-TSQR: Articulation of TSQR with QCG-OMPI}
\label{sec:contribution}

We explain in this section how we articulate the TSQR algorithm
with QCG-OMPI in order to confine intensive communications
within the different geographical sites of the computational grid.
The first difference from the TSQR algorithm as presented in
Section~\ref{sec:caqr} is that a domain is processed by a call to
ScaLAPACK (but not LAPACK as in~\cite{CAQR}). By doing so, we may attribute a domain to
a group of processes (instead of a single process) jointly performing
its factorization. The particular case of one domain per process
corresponds to the original TSQR (calls to LAPACK). At the other
end of the spectrum, we may associate one domain per geographical
site of the computational grid. The choice of the number of domains
impacts performance, as we will illustrate in Section~\ref{sec:xp-tsqr}. In all
cases, we call our algorithm TSQR (or QCG-TSQR), since it is a single
reduce operation based on a binary tree, similarly to the algorithm
presented in Section~\ref{sec:caqr}.

As explained in Section~\ref{sec:qcg-ompi}, the first task of
developing a QCG-OMPI application consists of defining the kind of
topologies expected by the application in a JobProfile. To get enough
flexibility, we request that processes are split into groups of
equivalent computing power, with good network connectivity inside
each group (low latency, high bandwidth) and we accept a lower network
connectivity between the groups. This corresponds to the classical cluster of
clusters approach, with a constraint on the relative size of the
clusters to facilitate load balancing.

The meta-scheduler will allocate resources in the physical grid that
matches these requirements. To enable us to complete an exhaustive
study on the different kind of topologies we can get, we also
introduced more constraints in the reservation mechanism, depending on
the experiment we ran. For each experiment, the set of machines that
are allocated to the job are passed to the MPI middleware, which
exposes those groups using two-dimensional arrays of group identifiers
(the group identifiers are defined in the JobProfile by the
developer). 
After the initialization, the application retrieves these group
identifiers from the system (using a specific MPI attribute) and
then creates one MPI communicator per group, using the
\emph{MPI\_Comm\_split} routine. Once this is done, the TSQR algorithm
has knowledge of the topology that allows it to adapt to the physical
setup of the grid.

The choice to introduce a requirement of similar computing power
between the groups however introduces constraints on the reservation
mechanism. For example, in some experiments discussed later
(Section~\ref{sec:xp}), only half the cores of some of the machines
were allocated in order to fit this requirement. Another possibility
would have been to handle load balancing issues at the algorithmic
level (and not at the middleware level) in order to relieve this
constraint on the JobProfile and thus increase the number of physical
setups that would match our needs. In the particular case of TSQR,
this is a natural extension; we would only have to adapt the number of
rows attributed to each domain as a function of the processing power 
dedicated to a domain. This alternative approach is future work.

\section{Performance model}
\label{sec:model}

\label{sec:opcount-stability}

In Tables~\ref{table:Ronly Flops} and~\ref{table:QR Flops}, we give the amount
of communication and computation required for ScaLAPACK QR2 and TSQR in two
different scenarios: first, when only the R-factor is requested
(Table~\ref{table:Ronly Flops}) and, second, when both the R-factor and the
Q-factor are requested (Table~\ref{table:QR Flops}). In this model, we assume
that a binary tree is used for the reductions and a homegeneous network.
We recall that the input matrix $A$ is $M$--by--$N$ and that $P$ is the number of domains.
The number of FLOPS is the number of FLOPS on the critical path per domain.

\begin{table}
\centering
\begin{tabular}{|c||c|c|c|}
\hline
& \# msg & vol. data exchanged & \# FLOPs \\
\hline
ScaLAPACK QR2  &  $ 2\mathbf{N}\log_2(P) $ & $\log_2(P) (N^2/2) $ & $(2MN^2 -2/3N^3)/P$             \\
TSQR   &  $ \log_2(P)  $ & $\log_2(P) (N^2/2) $ & $(2MN^2 -2/3N^3)/P+\mathbf {2/3\log_2(P)N^3}$   \\
\hline
\end{tabular}
\caption{\label{table:Ronly Flops}
Communication and computation breakdown when only the R-factor is needed.
}
\end{table}


Assuming a homogeneous network,
the total time of the factorization is then approximated by the formula:
\begin{equation}\label{eq:time}
\textmd{time } = \beta *  \textmd{ (\# msg) }  +\alpha *\textmd{ (vol. data exchanged) }+\gamma* \textmd{ (\# FLOPs) }, 
\end{equation}
where
 $\alpha$ is
 the inverse of the bandwidth,
 $\beta$
 the latency,
 and $\gamma$ the inverse of the floating point rate of a domain.
Although this model is simplistic, it enables us to forecast the basic trends. Note that in the case of TS matrices, 
we have $M\gg N$.

First we observe that the cost to compute both the $Q$ and the $R$ factors is exactly twice the cost for computing $R$ only.
Moreover, further theoretical and experimental analysis of the algorithm (see~\cite{CAQR}) reveal that the structure of the computation is the same
in both cases and the time to obtain $Q$ is twice the time to obtain $R$.
This leads to Property~\ref{prop:QandRistwiceR}.
For brevity, we mainly focus our study on the computation of $R$ only.
\begin{property}
\label{prop:QandRistwiceR}
The time to compute both $Q$ and $R$ is about twice the cost for computing $R$ only.
\end{property} 

One of the building blocks of the ScaLAPACK PDGEQR2 implementation and of our TSQR algorithm is the domanial QR factorization of a
TS matrix. The domain can be processed by a core, a node or a group of nodes.
We can not expect performance from our parallel distributed algorithms
to be better than the one of its domanial kernels.
This leads to Property~\ref{prop:low}.
In practice, the performance of the QR factorization of TS
matrices obtained from LAPACK/ScaLAPACK on a domain (core, node, small number of
nodes) is a small fraction of the peak.
(Term $\gamma$ of Equation~\ref{eq:time} is likely to be small.)
\begin{property}
\label{prop:low}
The performance of the factorization of TS matrices is
limited by the domanial performance of the QR factorization of TS matrices.
\end{property}

We see that the number of operations is proportional to $m$ while all the
communication terms (latency and bandwidth) are independent of $m$.  Therefore
when $m$ increases, the communication time stays constant whereas the domanial
computation time increases. This leads to increased performance.
\begin{property}
 \label{prop:m}
The performance of the factorization of TS matrices
increases with $M$.
\end{property} 

The number of operations is proportional to $N^2$ while the number of messages is proportional to $N$.
Therefore when $N$ increases, the latency term is hidden by the computation term. This leads to better performance. We also note that
increasing $N$ enables better performance of the domanial kernel since
it can use Level 3 BLAS when 
the number of columns is greater than, perhaps, $100$.
This is Property~\ref{prop:n}.
\begin{property}
 \label{prop:n}
The performance of the factorization of TS matrices
increases with $N$.
\end{property} 

Finally, we see that the latency term is $2\log_2(P)$ for TSQR while it is
$2\mathbf{N} \log_2(P)$ for ScaLAPACK QR2. On the other hand, the FLOPs term
has a non parallelizable additional $\mathbf{2/3\log_2(P)N^3}$ term for the
TSQR algorithm.  We see that TSQR effectively trades messages for flops. We
expect TSQR to be faster than ScaLAPACK QR2 for $N$ in the mid-range (perhaps
between five and a few hundreds). For larger $N$, TSQR will become slower because
of the additional flops. This is Property~\ref{prop:tsqrbetter}. (We note that for
large $N$, one should stop using TSQR and switch to CAQR.)

\begin{table}
\centering
\begin{tabular}{|c||c|c|c|}
\hline
 & \# msg & vol. data exchanged & \# FLOPs  \\
\hline
ScaLAPACK QR2   &  $ 4\mathbf{N}\log_2(P) $ & $2\log_2(P) (N^2/2) $ & $(4MN^2 -4/3N^3)/P$             \\
TSQR   &  $ 2\log_2(P)  $ & $2\log_2(P) (N^2/2) $ & $(4MN^2 -4/3N^3)/P+\mathbf{4/3\log_2(P)N^3}$   \\
\hline
\end{tabular}
\caption{\label{table:QR Flops}
Communication and computation breakdown when both the Q-factor and the R-factor are needed.
}
\end{table}

\begin{property}
 \label{prop:tsqrbetter}
The performance of TSQR is better than ScaLAPACK for $N$ in the mid range.
When $N$ gets too large, the performance of TSQR deteriorates and ScaLAPACK becomes better.
\end{property} 

\newdimen\figurewidth
\figurewidth=0.35\linewidth

\section{Experimental study}
\label{sec:xp}

\subsection{Experimental environment}
\label{sec:xp-env}

We present an experimental study of the performance of the QR
factorization of TS matrices in a grid computing environment. We
conducted our experiments on Grid'5000. This platform is a dedicated,
reconfigurable and controllable experimental grid of 13 clusters
distributed over 9 cities in France. Each cluster is itself composed
of 58 to 342 nodes. The clusters are inter-connected through dedicated
black fiber. In total, Grid'5000 roughly gathers $5,000$ CPU cores featuring 
multiple architectures.

For the experiments presented in this study, we chose four clusters
based on relatively homogeneous dual-processor nodes, ranging from AMD
Opteron 246 (2 GHz/1MB L2 cache) for the slowest ones to AMD Opteron
2218 (2.6 GHz/2MB L2 cache) for the fastest ones, which leads to
theoretical peaks ranging from $8.0$ to $10.4$~Gflop/s per processor. 
These four clusters are the 93-node cluster in Bordeaux,  
the 312-node cluster in Orsay,  
a 80-node cluster in Toulouse,  
and a 56-node cluster in Sophia-Antipolis.  
Because these clusters are located in different
cities, we will indistinctly use the terms \emph{cluster} and
\emph{geographical site} (or \emph{site}) in the following. Nodes are
interconnected with a Gigabit Ethernet switch; on each node, the
network controller is shared by both processors. On each cluster, we
reserved a subset of 32 dual-processor nodes, leading to a
theoretical peak of $512.0$ to $665.6$~Gflop/s per node. Our algorithm
being synchronous, to evaluate the proportion of theoretical peak achieved
in an heterogeneous environment, we consider the efficiency of the slowest
component as a base for the evaluation. Therefore, the theoretical peak of our grid is
equal to $2,048$~Gflop/s. A consequence of the constraints on the topology
expressed by our implementation in QCG-OMPI (see Section~\ref{sec:qcg-ompi})
is that in some experiments,
machines with dual 2-cores processors were booked with the ability to use 2 cores
(over 4) only.
\ignore{In some of the experiments
(Section~\ref{sec:xp-dedicated}), we turned off one processor per
node to emulate a grid with (i) a different ratio computational power
- throughput (ii) a network controller dedicated to one processor
(thus avoiding concurrent accesses to the controllers). In this latter
configuration, the theoretical peaks range from $256.0$ to
$332.8$~Gflop/s for a single node and thus a total $1,024$~Gflop/s for
that simulated grid.}

The performance of the inter and intra-cluster communications is shown
in Table~\ref{tab:comm}. Within a cluster, nodes are connected with a
GigaEthernet network. Clusters are interconnected with $10$ Gb/s dark
fibers. The intra-cluster throughput is consistently equal to $890$
Mb/s but varies from $61$ to $860$ Mb/s between
clusters. Inter-cluster latency is roughly greater than intra-cluster
latency by two orders of magnitude. Between two processors of a same
node, OpenMPI uses a driver optimized for shared-memory architectures,
leading to a $17$ $\mu{}$s latency and a $5$ Gb/s throughput.

One major feature of the Grid5000 project is the ability of the user
to boot her own environment (including the operating system,
distribution, libraries, etc.) on all the computing nodes booked for
her job. All the nodes were booted under Linux 2.6.30. The tests and
benchmarks were compiled with GCC 4.0.3 (flag -O3) and run in
dedicated mode (no other user can access the machines). ScaLAPACK
1.8.0 and GotoBLAS 1.26 libraries were used. Finally we recall that we focus on the
factorization of TS dense large-scale matrices in real double
precision, corresponding to up to $16$~GB of memory (\emph{e.g.} a
$33,554,432 \times 64$ matrix in double precision).


\subsection{Tuning of the applications}
\label{sec:xp-tuning}

\ignore{
\begin{figure}[htb]
  \centering
  \begin{tabular}{cc}
    \begin{minipage}{\figurewidth}
      \includegraphics[angle=270,width=\linewidth]{dgemm}
      \caption{DGEMM (matrix multiplication) performance on a
        dual-processor AMD Opteron 246 node at Orsay.}
      \label{fig:dgemm}
    \end{minipage}
    &
    \begin{minipage}{\figurewidth}
      \includegraphics[angle=270,width=\linewidth]{sousdomaines}
      \caption{Effect of the number of domains on TSQR performance.
        Single site (32 nodes at Orsay).}
      \label{fig:sousdomaines}
      
    \end{minipage}
    \\
  \end{tabular}
\end{figure}
}

To achieve high performance across platforms, dense linear algebra
applications rely on Basic Linear Algebra Subprograms
(BLAS)~\cite{blas} to perform basic operations such as vector and
matrix multiplication. This design by layers allows one to only focus
on the optimization of these basic operations while keeping underlying
numerical algorithms common to all machines. From the performance of
BLAS operations -- and in particular the matrix multiplication (DGEMM)
-- thus depends the behavior of the overall
application. The Automatically Tuned Linear Algebra Software
(ATLAS)~\cite{atlas} library is, for instance, a widely used
implementation of BLAS achieving high performance thanks to autotuning
methods. It is furthermore possible to take advantage of
dual-processor nodes thanks to a multi-threaded implementation of BLAS
such as GotoBLAS~\cite{gotoblas}. We have compared the performance of
serial and multi-threaded GotoBLAS DGEMM against ATLAS. Both
configurations of GotoBLAS outperformed ATLAS; we thus selected GotoBLAS
to conduct our experiments. Another possibility to take advantage of
dual-processor nodes is simply to create two processes per node at the
application level. For both ScaLAPACK and TSQR, that latter
configuration consistently achieved a higher performance. \ignore{(we
  valid this claim in Section~\ref{sec:xp-dedicated}). Except in
  Section~\ref{sec:xp-dedicated} where we emulate special
  configurations, w} We therefore used \emph{two processes per node
  together with the serial version of GotoBLAS's DGEMM in all the
  experiments} reported in this study. With DGEMM being the fastest kernel
(on top of which other BLAS operations are usually built), we obtain a
rough practical performance upper bound for our computational grid of
about $940$~Gflop/s (the ideal case where 256 processors would achieve
the performance of DGEMM, \emph{i.e.}, about $3.67$~Gflop/s each) out
of the $2,048$~Gflop/s theoretical peak.

SCALAPACK implements block-partitioned algorithms. Its performance
depends on the partitioning of the matrix into blocks. Preliminary
experiments (not reported here) showed that a column-wise 1D-cyclic
partition is optimum for processing TS matrices in our environment.
We furthermore chose a block size consistently equal to $64$ (a better
tuning of this parameter as a function of the matrix characteristics
would have occasionally improved the performance but we considered
that the possible gain was not worth the degree of complexity
introduced in the analysis of the results.).

\subsection{ScaLAPACK performance}
\label{sec:xp-scalapack}

\begin{figure}[htbp]
  \centering
  \subfigure[N = 64]{
    \includegraphics[angle=270,width=\figurewidth]{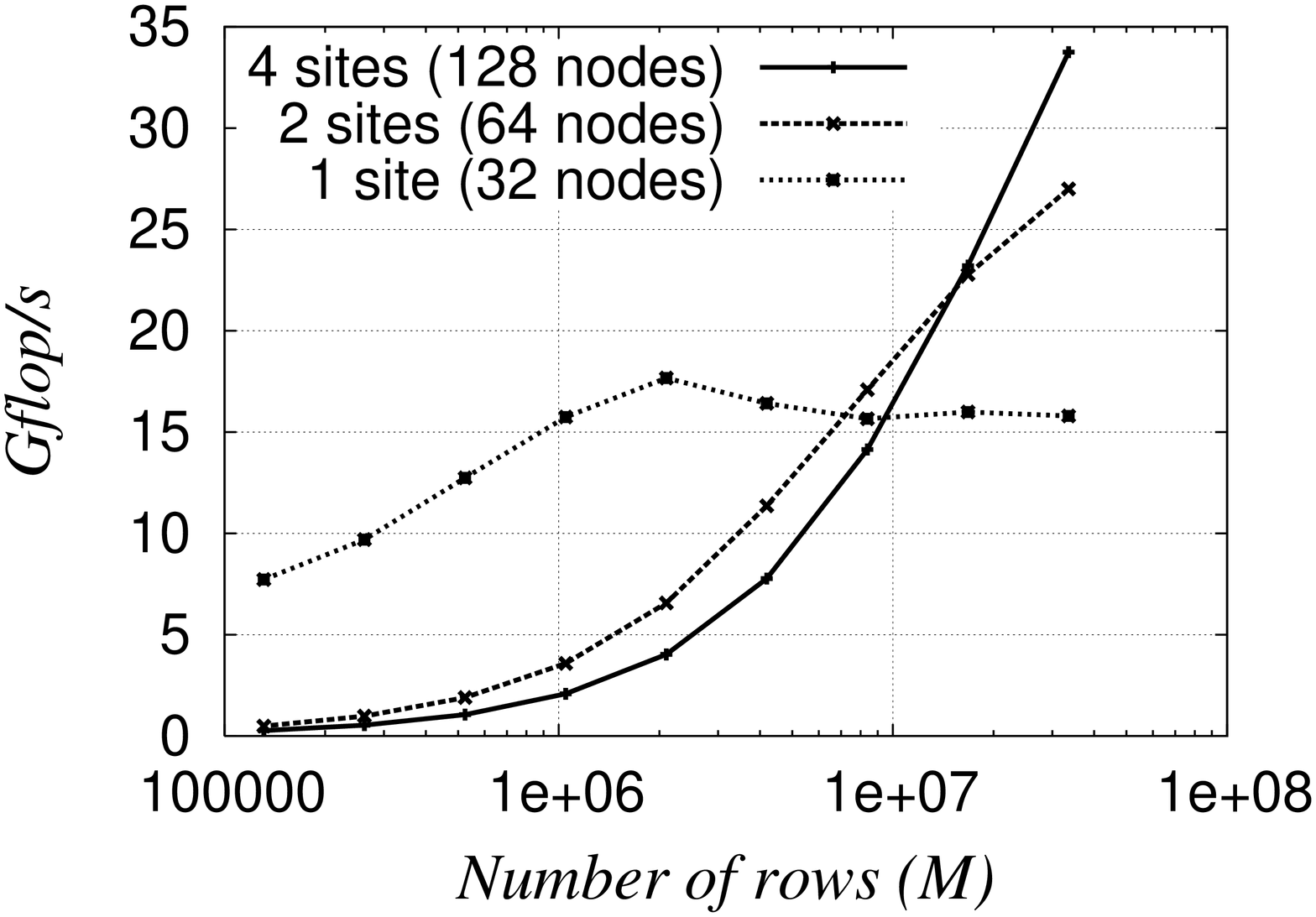}
    \label{fig:QR_nb1}
  }
  \subfigure[N = 128]{
    \includegraphics[angle=270,width=\figurewidth]{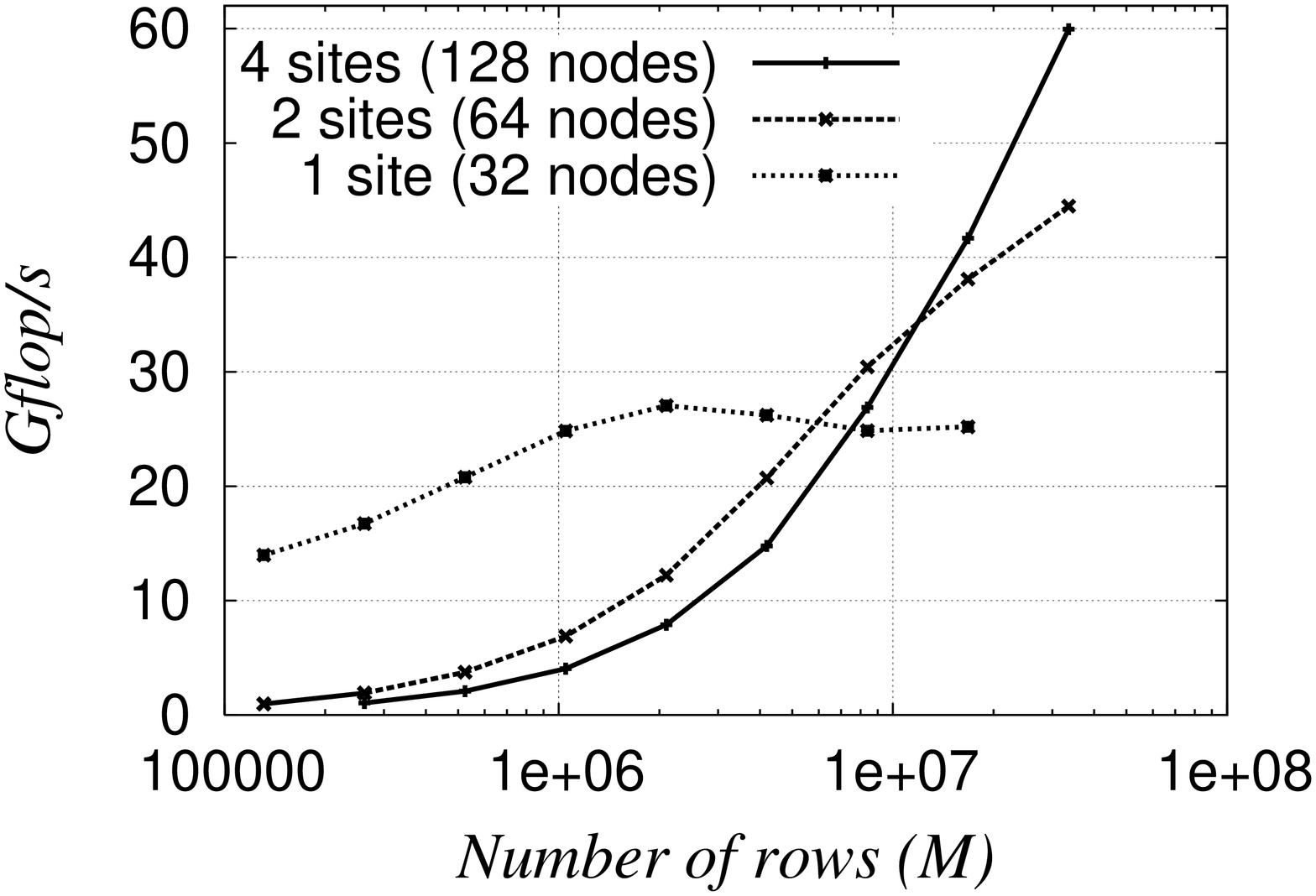}
    \label{fig:QR_nb2}
  }\\
  \subfigure[N = 256]{
    \includegraphics[angle=270,width=\figurewidth]{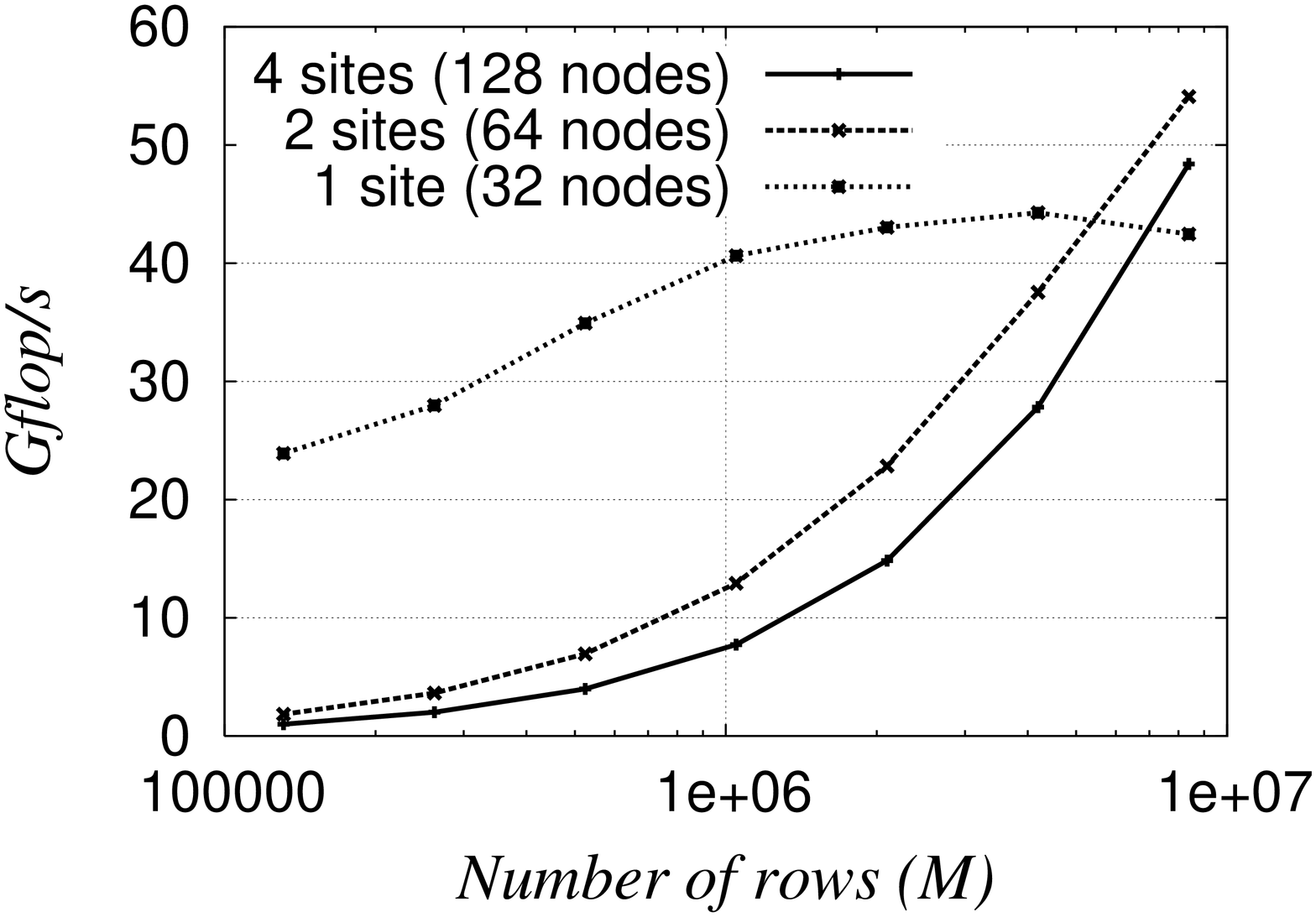}
    \label{fig:QR_nb4}
  }
  \subfigure[N = 512]{
    \includegraphics[angle=270,width=\figurewidth]{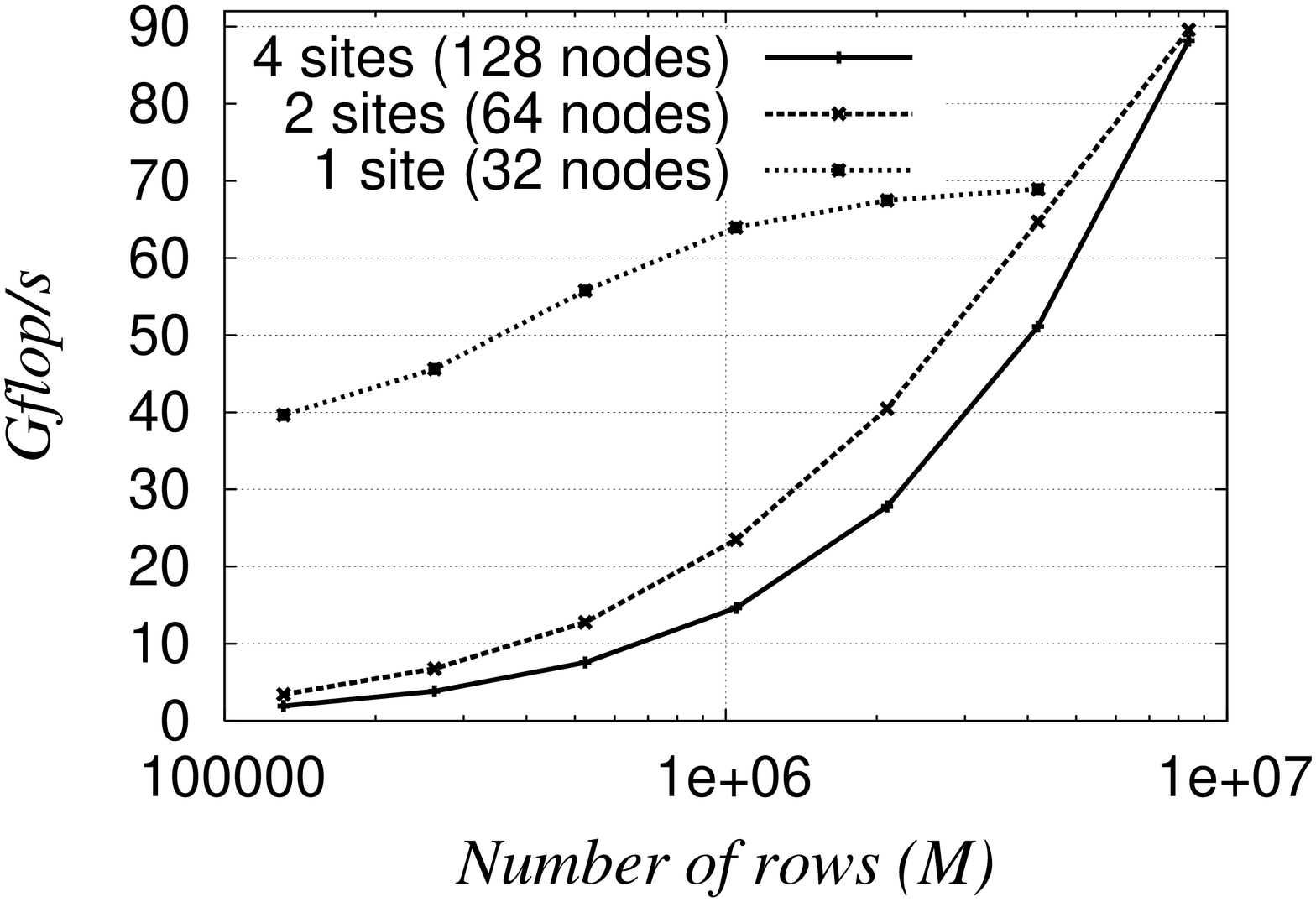}
    \label{fig:QR_nb8}
  }
  \caption{ScaLAPACK performance.}
  \label{fig:QR-2p1t}
\end{figure}

Figure~\ref{fig:QR-2p1t} reports ScaLAPACK performance. In accordance
with Property~\ref{prop:low}, the overall performance of the QR
factorization of TS matrices is low (consistently lower than
$90$~Gflop/s) compared to the practical upper bound of our grid
($940$~Gflop/s). Even on a single cluster, this ratio is low since the
performance at one site is consistently lower than $70$~Gflop/s out of
a practical upper bound of $235$~Gflop/s. As expected too
(properties~\ref{prop:m} and~\ref{prop:n}), the performance increases
with the dimensions of the matrix. For matrices of small to moderate
height ($M \leq 5,000,000$), the fastest execution is consistently the
one conducted on a single site. In other words, for those matrices,
the use of a grid (two or four sites) induces a drop in 
performance, confirming previous
studies~\cite{grads,gridtop500,Walker08}. For very tall matrices ($M >
5,000,000$), the proportion of computation relative to the amount of
communication becomes high enough so that the use of multiple sites
eventually speeds up the performance (right-most part of the graphs
and Property~\ref{prop:m}). This speed up however hardly surpasses a
value of $2.0$ while using four sites (Figure~\ref{fig:QR_nb2}).

\subsection{QCG-TSQR performance}
\label{sec:xp-tsqr}

\begin{figure}[htbp]
  \centering
  \subfigure[N = 64]{
    \includegraphics[angle=270,width=\figurewidth]{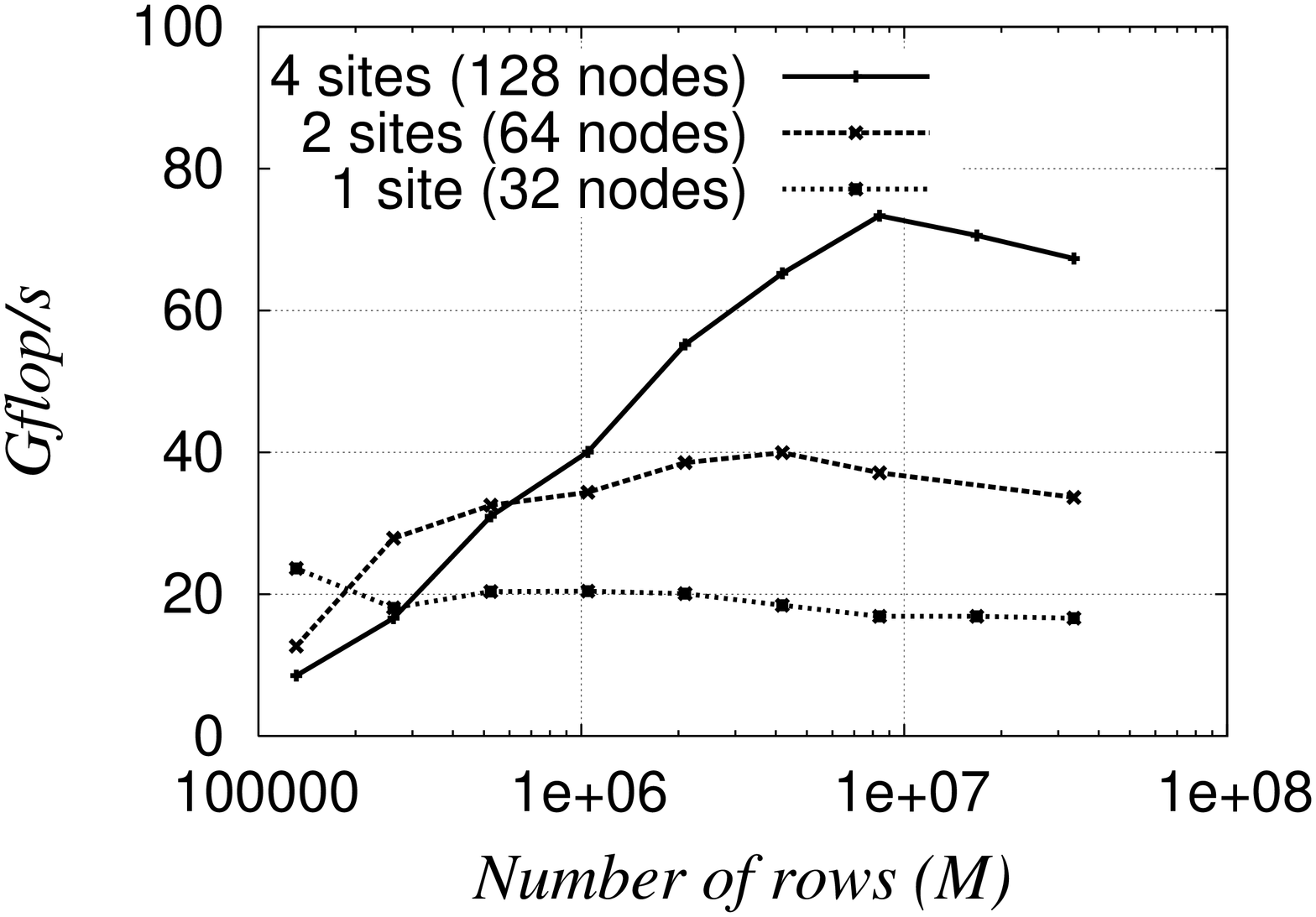}
    \label{fig:TSQR_nb1}
  }
  \subfigure[N = 128]{
    \includegraphics[angle=270,width=\figurewidth]{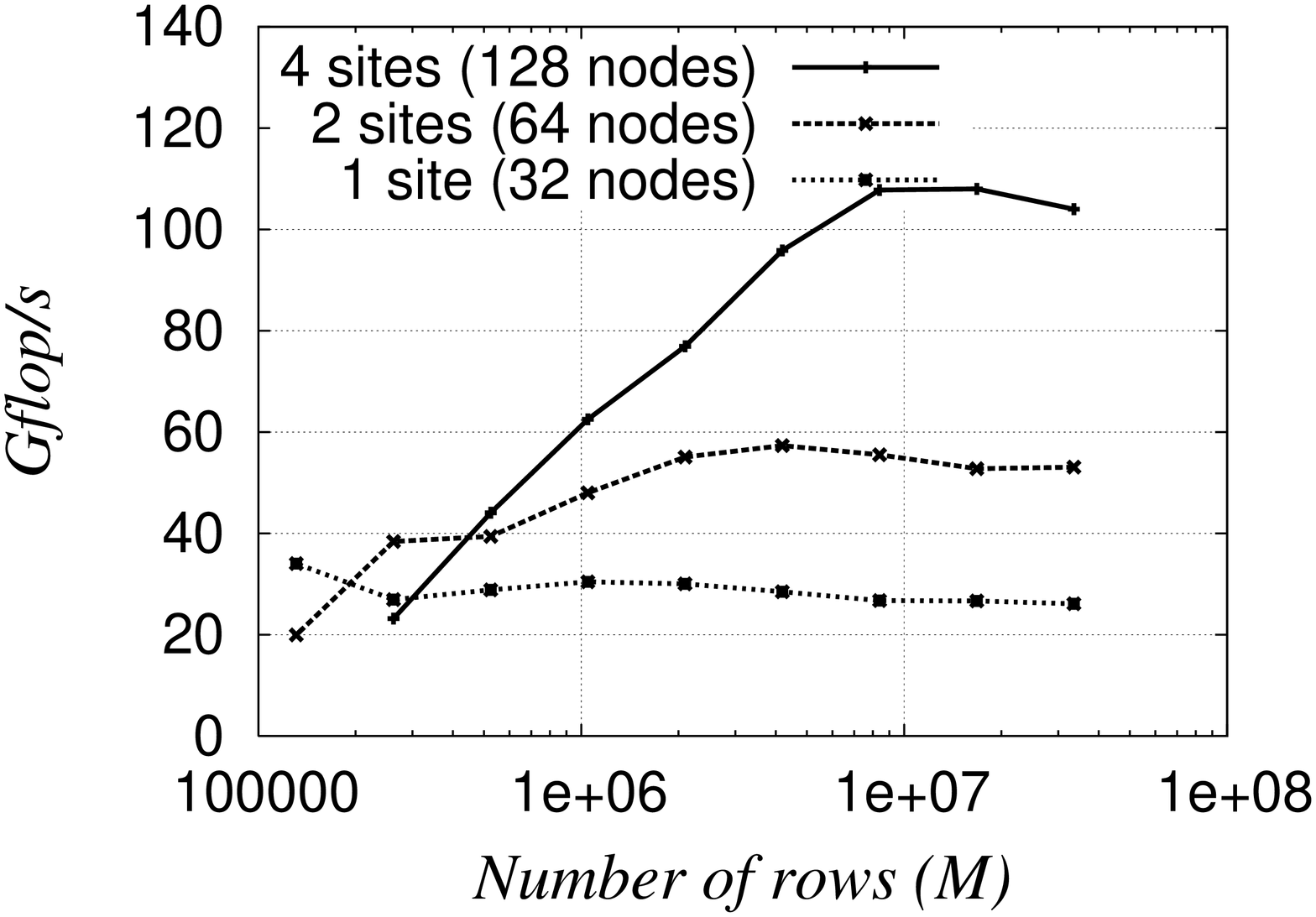}
    \label{fig:TSQR_nb2}
  }\\
  \subfigure[N = 256]{
    \includegraphics[angle=270,width=\figurewidth]{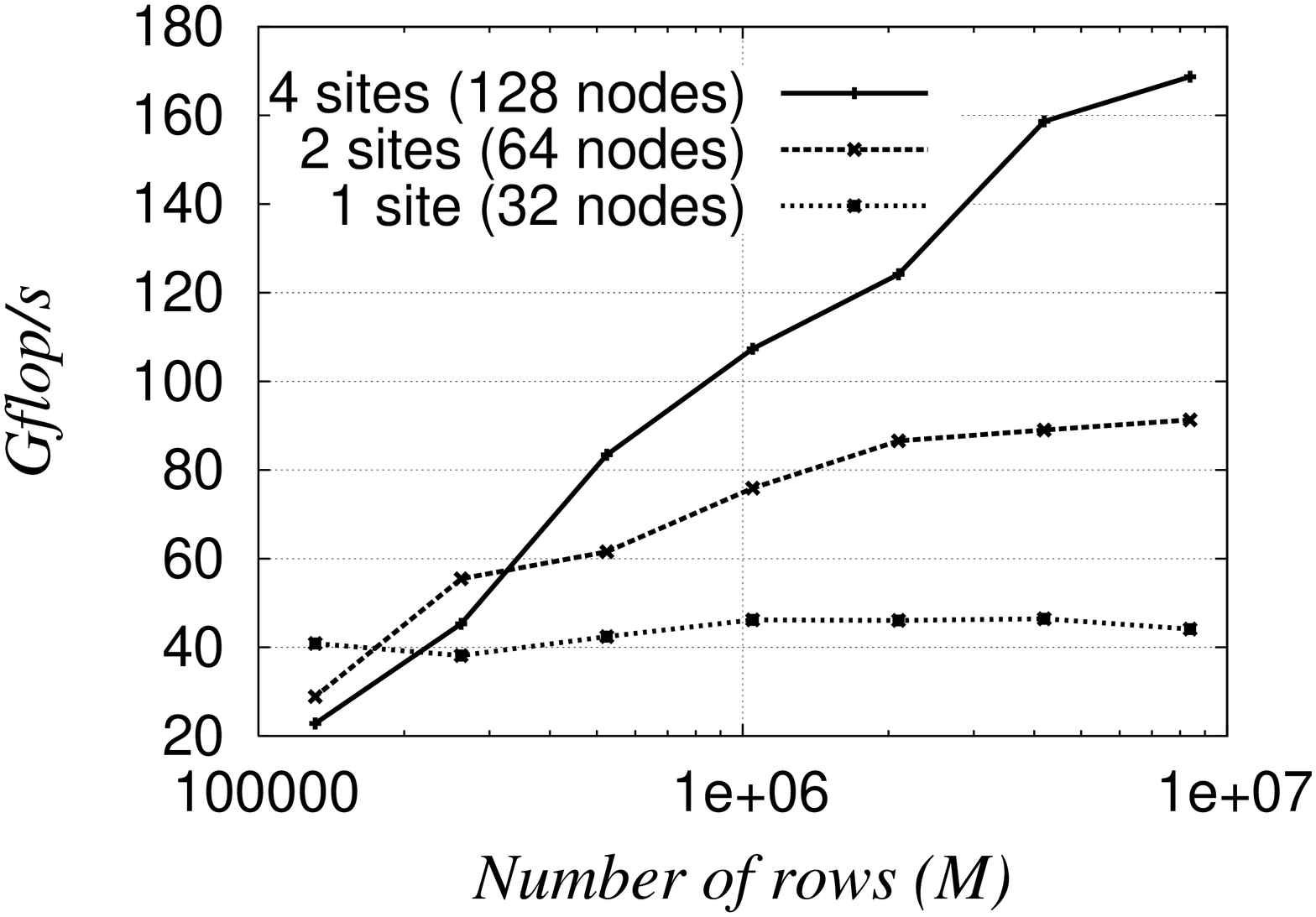}
    \label{fig:TSQR_nb4}
  }
  \subfigure[N = 512]{
    \includegraphics[angle=270,width=\figurewidth]{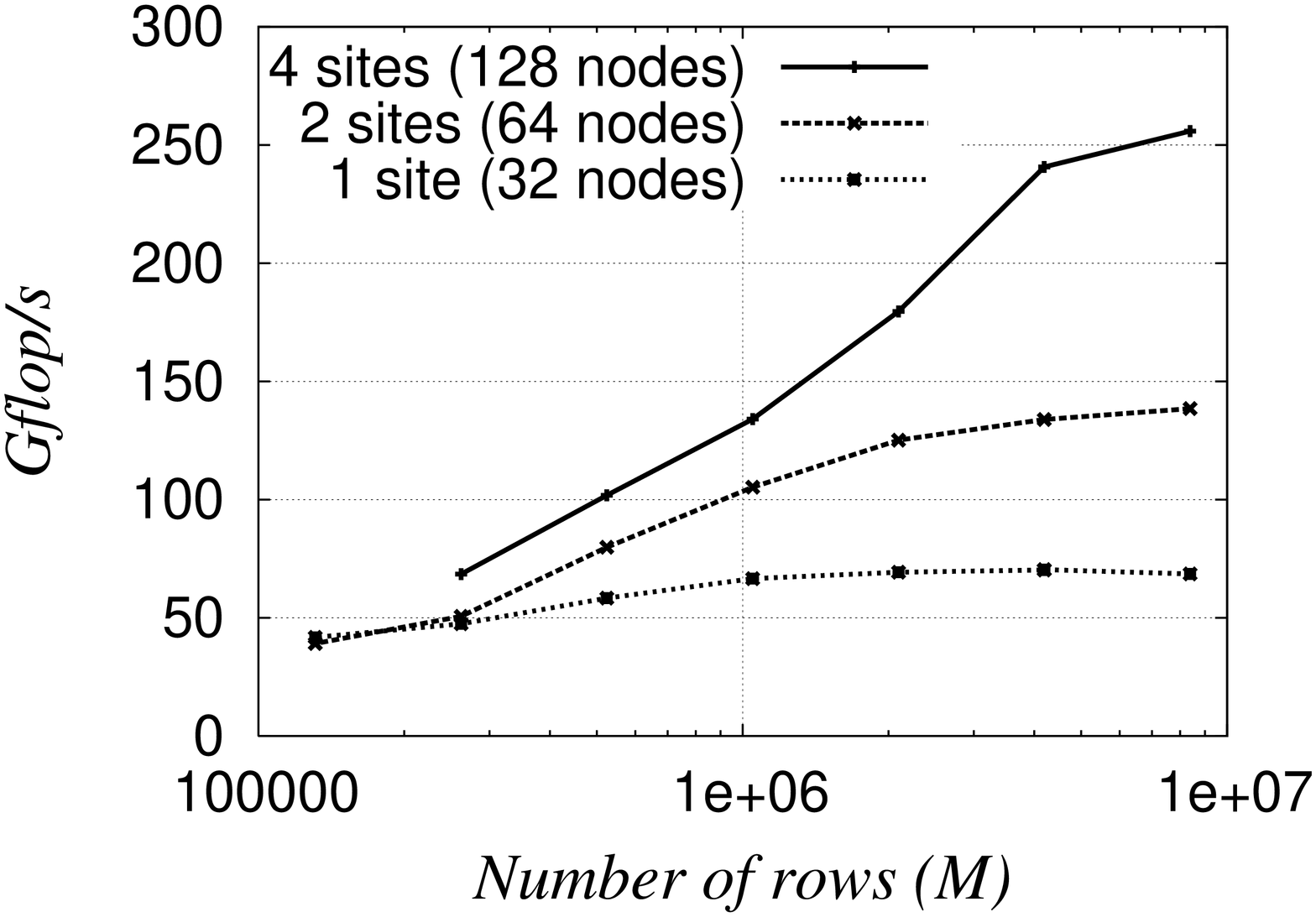}
    \label{fig:TSQR_nb8}
  }
  \caption{TSQR Performance.}
  \label{fig:TSQR-2p1t}
\end{figure}

The performance of TSQR (articulated with QCG-OMPI as described in
Section~\ref{sec:contribution}) depends on the number of domains
used. In Figure~\ref{fig:TSQR-2p1t}, we report the TSQR performance for
the optimum number of domains and we will return later to the
effect of the number of domains. In accordance with
Property~\ref{prop:low}, the overall performance is again only a
fraction of the practical upper bound of our grid ($940$~Gflop/s).
But, compared to ScaLAPACK, this ratio is significantly higher since
the factorization of a $8,388,608 \times 512$ matrix achieves $256$
Gflop/s (Figure~\ref{fig:TSQR_nb8}). Again, in accordance with
properties~\ref{prop:m} and~\ref{prop:n}, the overall performance
increases with the dimensions of the matrix. Thanks to its better
performance (Property~\ref{prop:tsqrbetter}), TSQR also achieves a speed up
on the grid on matrices of moderate size. Indeed, for almost all
matrices of moderate to great height ($M \geq 500,000$), the fastest
execution is the one conducted on all four sites. Furthermore, for very
tall matrices ($M \geq 5,000,000$), TSQR performance scales almost
linearly with the number of sites (a speed up of almost $4.0$ is
obtained on four sites). This result is the central statement of this
paper and validates the thesis that computational grids are a valid
infrastructure for solving large-scale problems relying on the QR
factorization of TS matrices.

\begin{figure}[htbp]
  \centering
  \subfigure[N = 64]{
    \includegraphics[angle=270,width=\figurewidth]{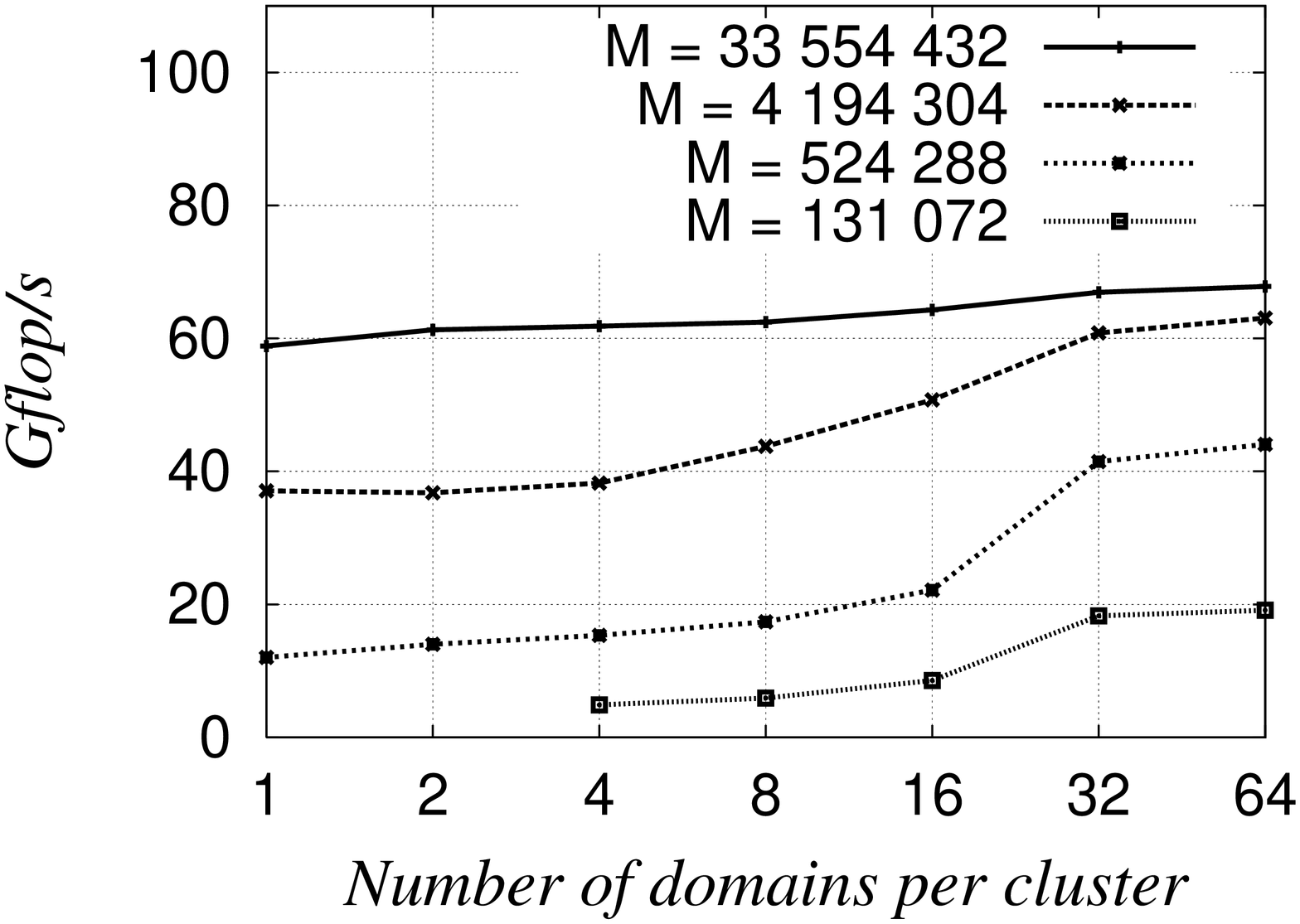}
    \label{fig:TSQR_domaines_nb1}
  }
  \subfigure[N = 128]{
    \includegraphics[angle=270,width=\figurewidth]{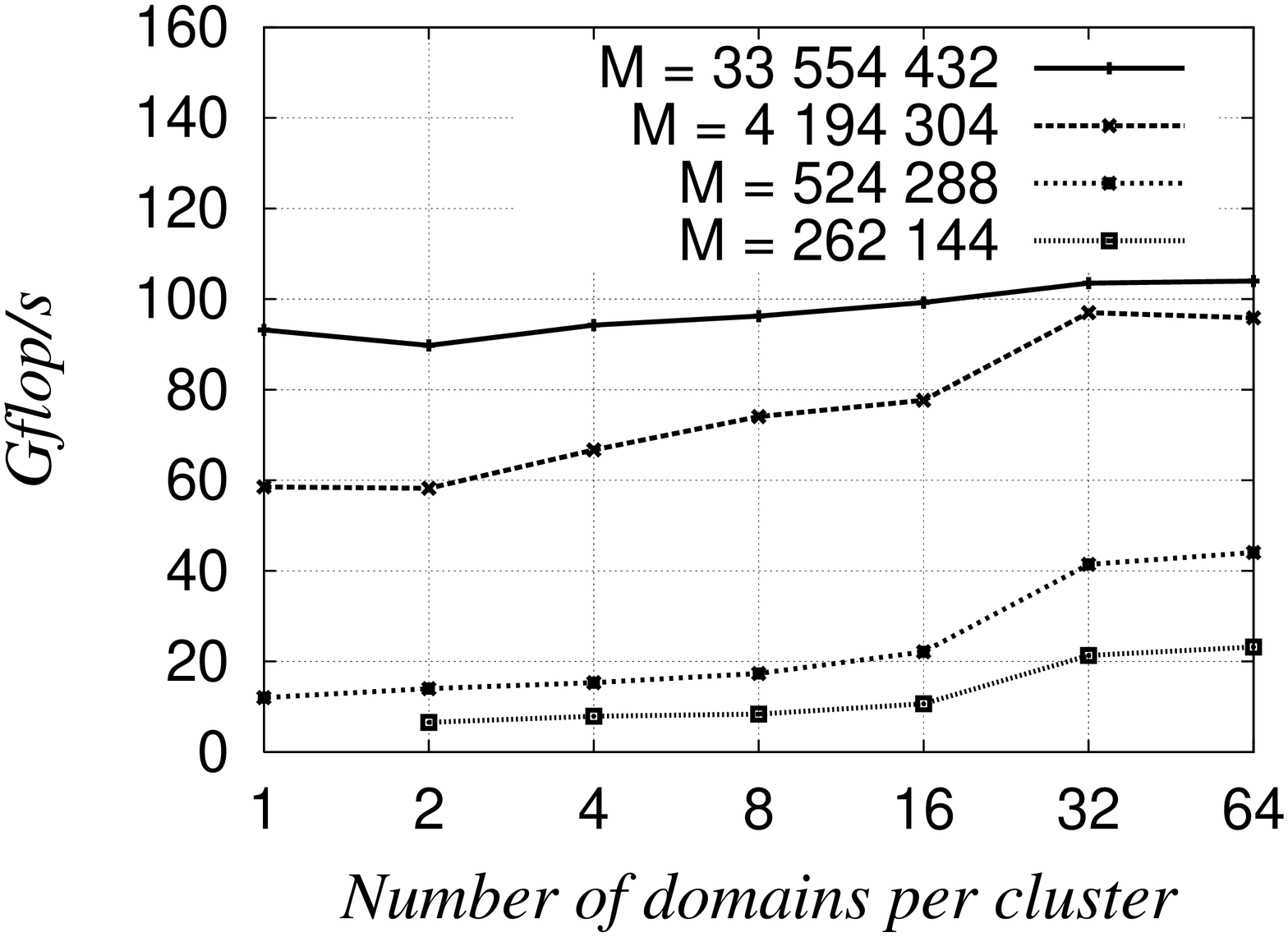}
    \label{fig:TSQR_domaines_nb2}
  }\\
  \subfigure[N = 256]{
    \includegraphics[angle=270,width=\figurewidth]{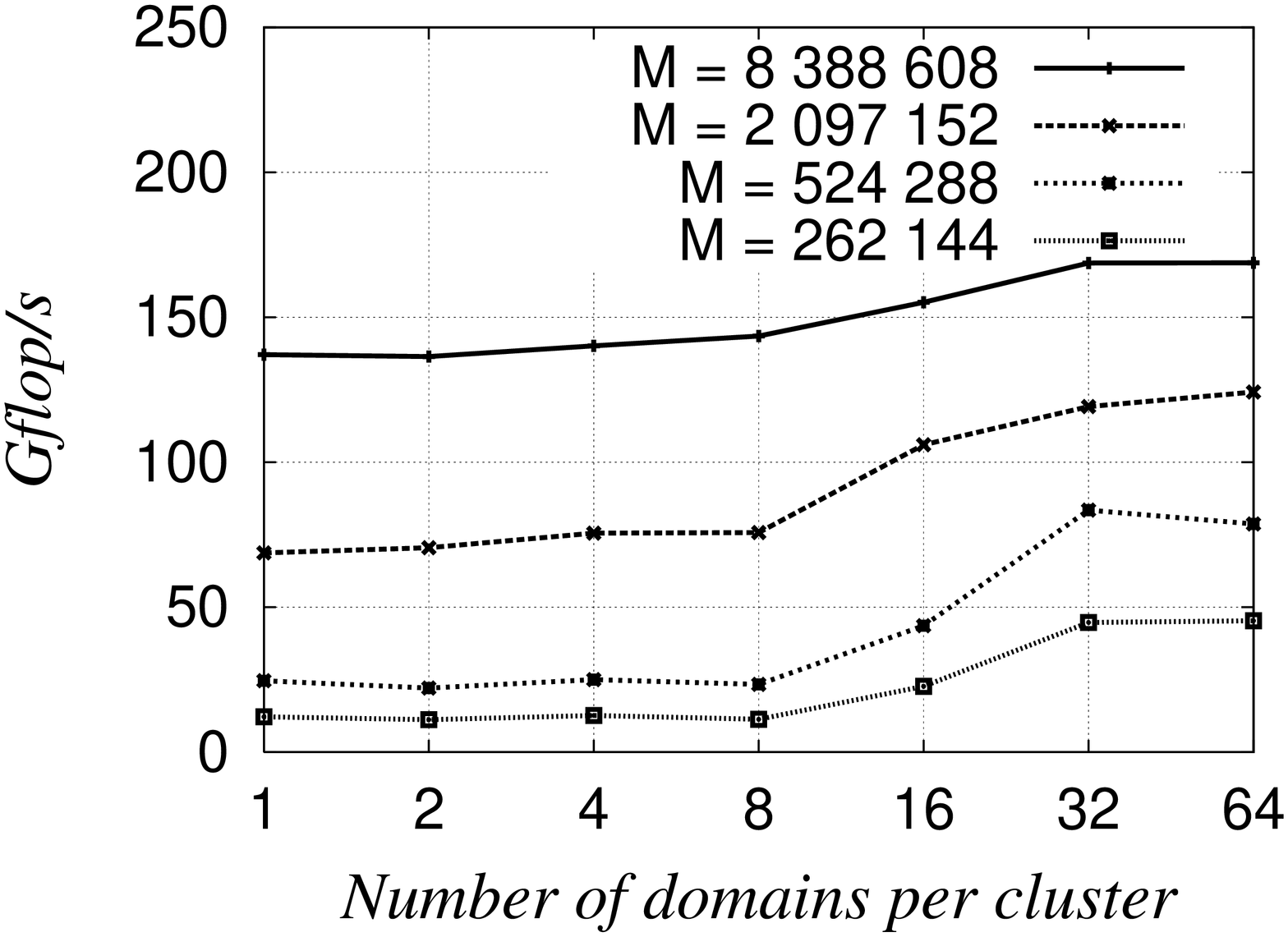}
    \label{fig:TSQR_domaines_nb4}
  }
  \subfigure[N = 512]{
    \includegraphics[angle=270,width=\figurewidth]{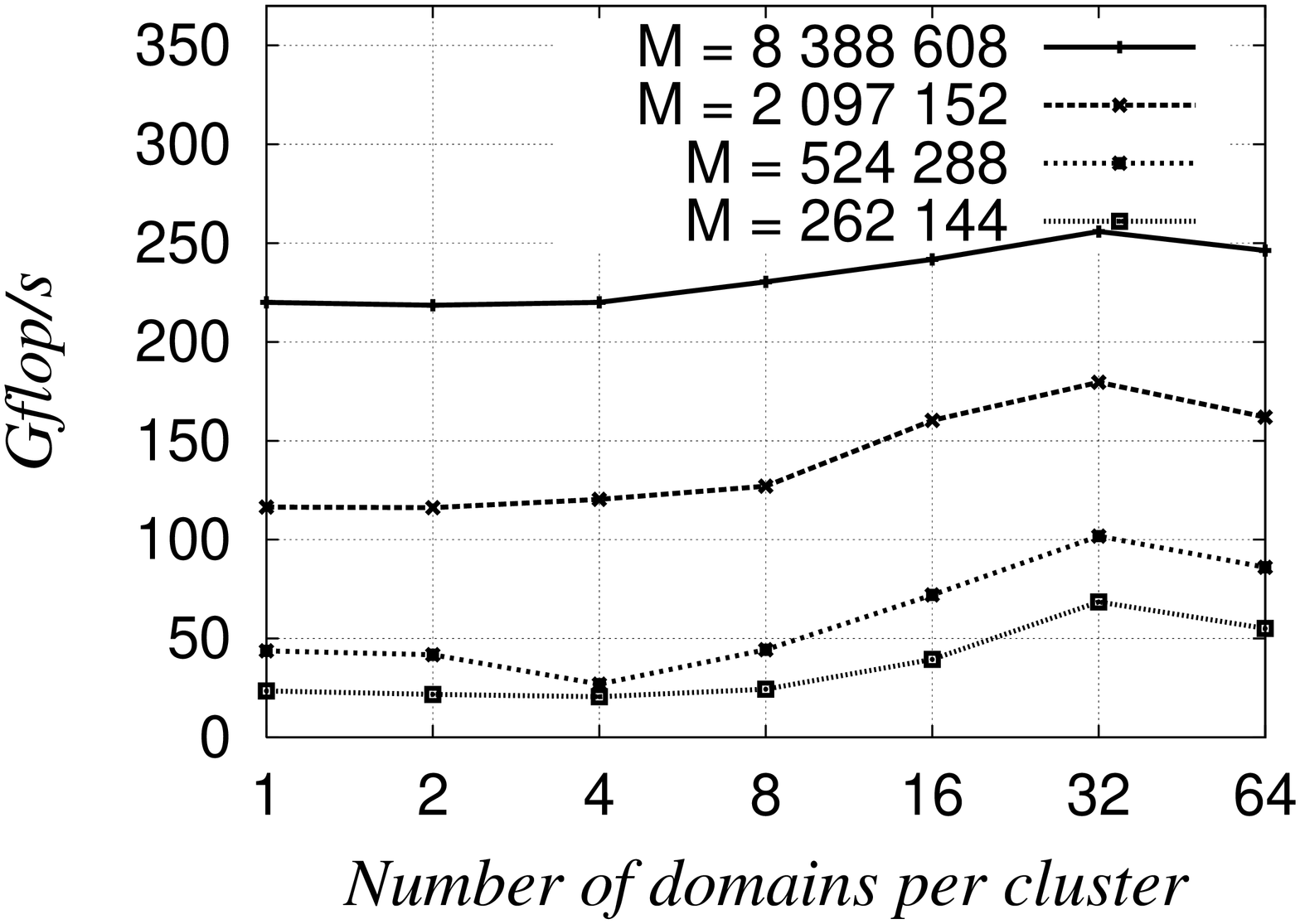}
    \label{fig:TSQR_domaines_nb8}
  }
  \caption{Effect of the number of domains on the performance of TSQR
    executed on all four sites.}
  \label{fig:TSQR_domaines_2p1t}
\end{figure}

Figure~\ref{fig:TSQR_domaines_2p1t} now illustrates the effect of the
number of domains per cluster on TSQR performance.
\begin{figure}[htbp]
  \centering
  \subfigure[N = 64]{
    \includegraphics[angle=270,width=\figurewidth]{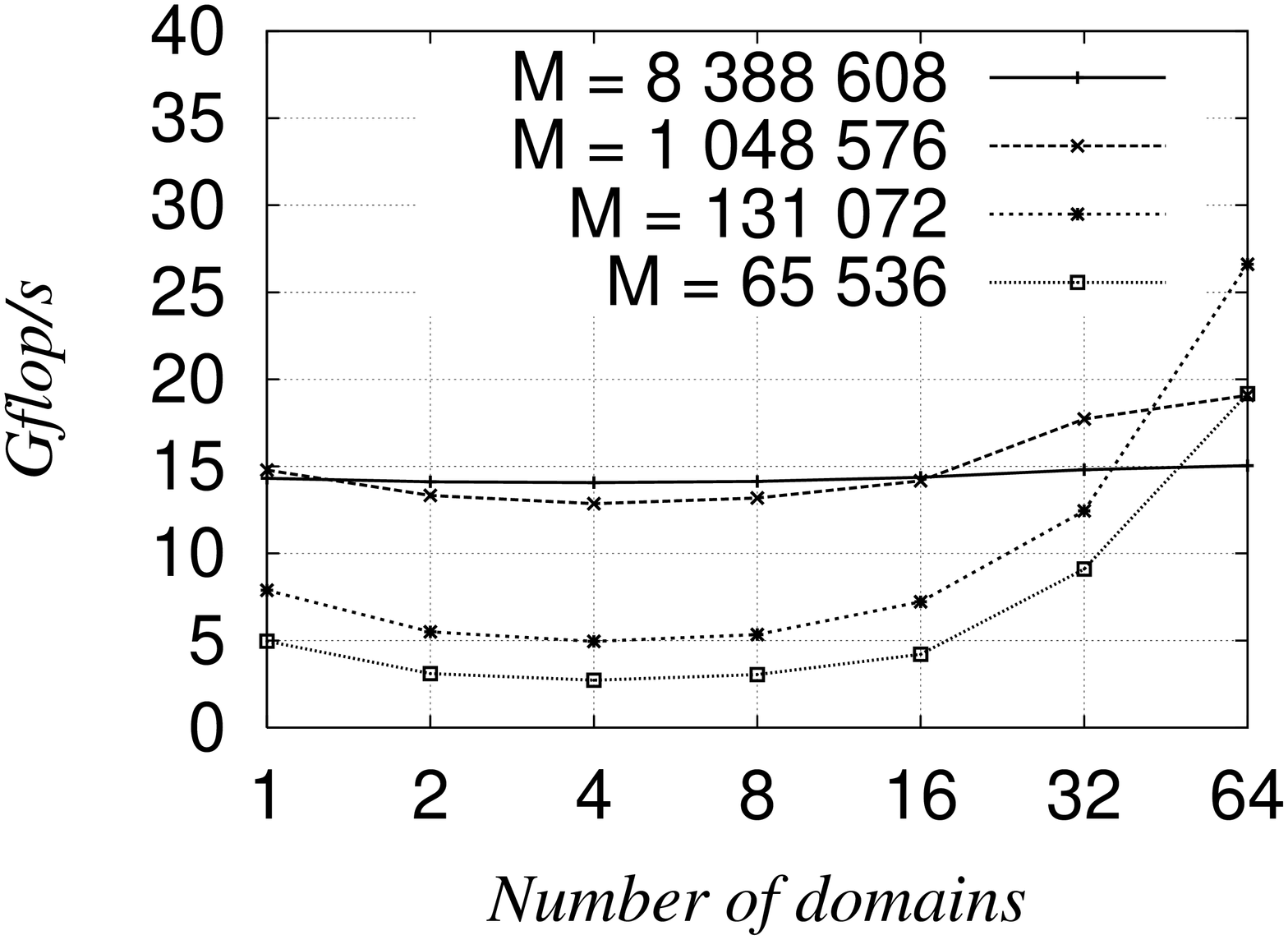}
    \label{fig:TSQR_1site_domaines_nb1}
  }
  \subfigure[N = 512]{
   \includegraphics[angle=270,width=\figurewidth]{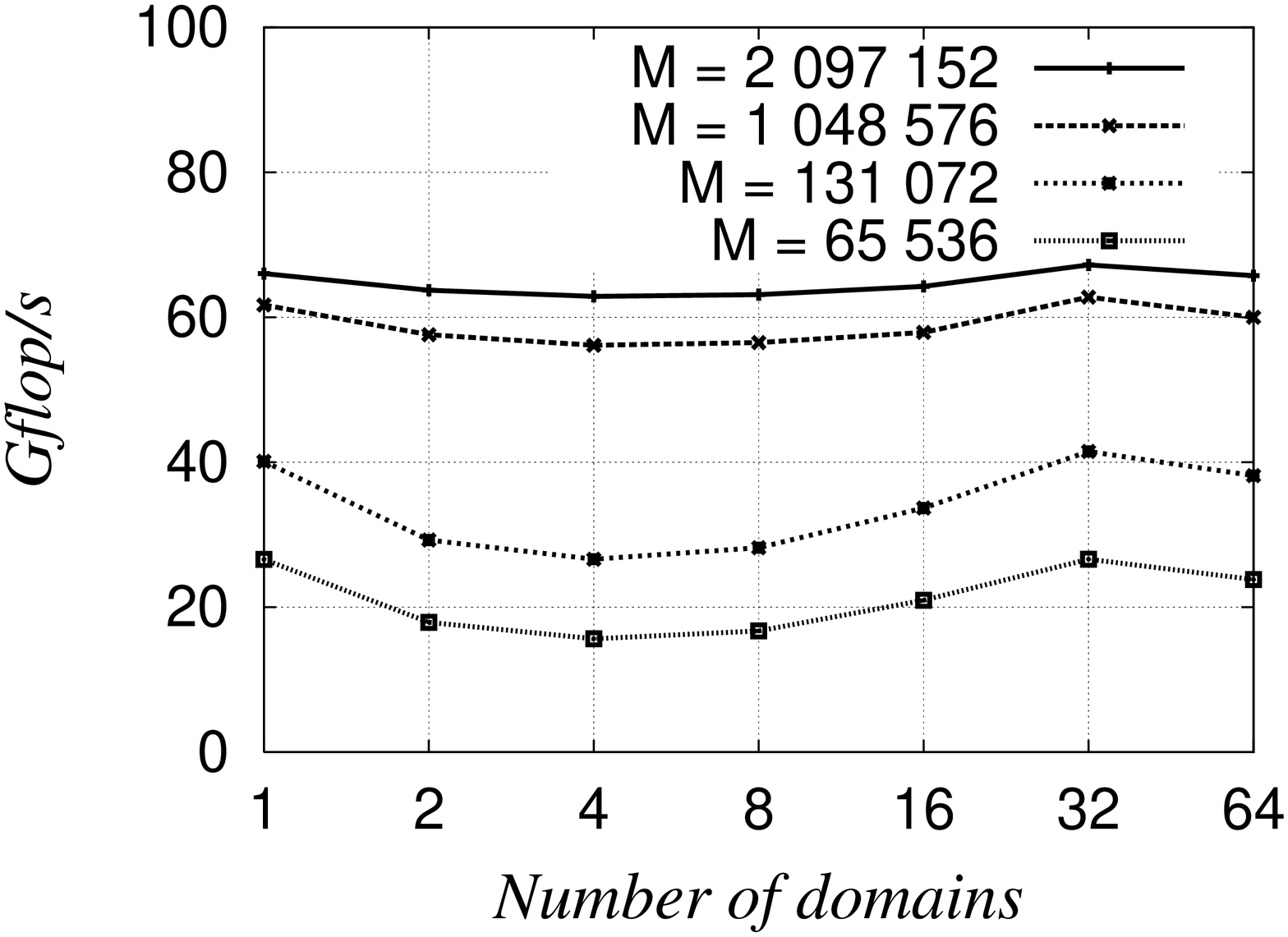}
   \label{fig:TSQR_1site_domaines_nb8}
  }
  \caption{Effect of the number of domains on the performance of TSQR
    executed on a single site.}
  \label{fig:TSQR_1site_domaines_2p1t}
\end{figure}
Globally, the performance increases with the number of domains. For
very tall matrices ($M = 33,554,432$), the impact is limited (but not
negligible) since there is enough computation to almost mask the
effect of communications (Property~\ref{prop:m}). For very skinny
matrices ($N=64$), the optimum number of domains for executing TSQR on
a single cluster is $64$ (Figure~\ref{fig:TSQR_1site_domaines_nb1}),
corresponding to a configuration with one domain per processor. This
optimum selection of the number of domains is translated to executions
on multiple clusters where $64$ domains per cluster is optimum too
(Figure~\ref{fig:TSQR_domaines_nb1}). For the widest matrices studied
($N=512$), the optimum number of domains for executing TSQR on a
single cluster is $32$ (Figure~\ref{fig:TSQR_1site_domaines_nb8}),
corresponding to a configuration with one domain per node. For those
matrices, trading flops for intra-node communications is not worthwhile.
This behavior is again transposable to executions on multiple sites 
(Figure~\ref{fig:TSQR_domaines_nb8}) where the optimum configuration
also corresponds to $32$ domains per cluster. This observation illustrates the fact that one should use CAQR
and not TSQR for large $N$, as discussed in Section~\ref{sec:model}.

\subsection{QCG-TSQR vs ScaLAPACK}
\label{sec:xp-vs}

\begin{figure}[htbp]
  \centering \subfigure[N = 64]{
    \includegraphics[angle=270,width=\figurewidth]{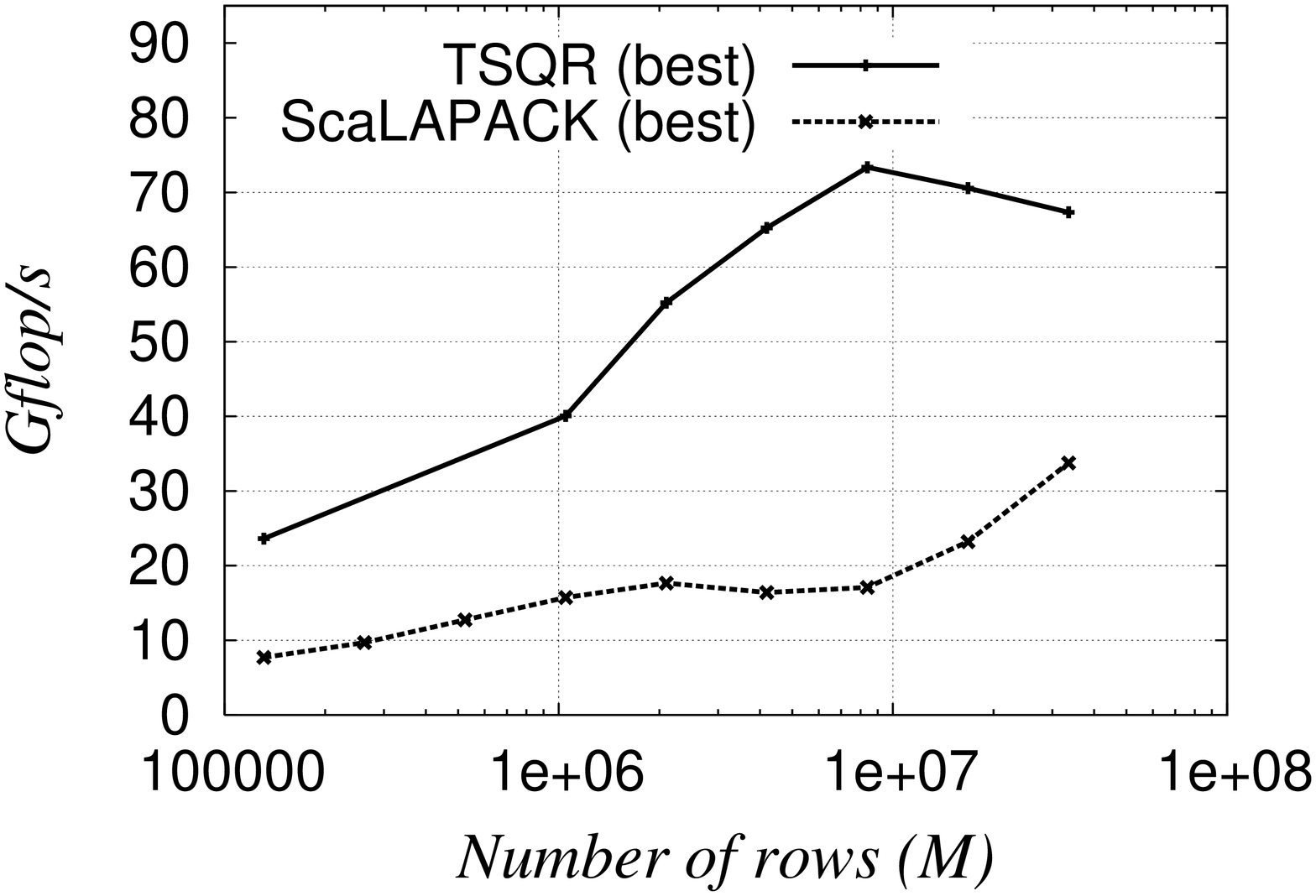}
    \label{fig:TSQR_QR_nb1}
  }
  \subfigure[N = 128]{
    \includegraphics[angle=270,width=\figurewidth]{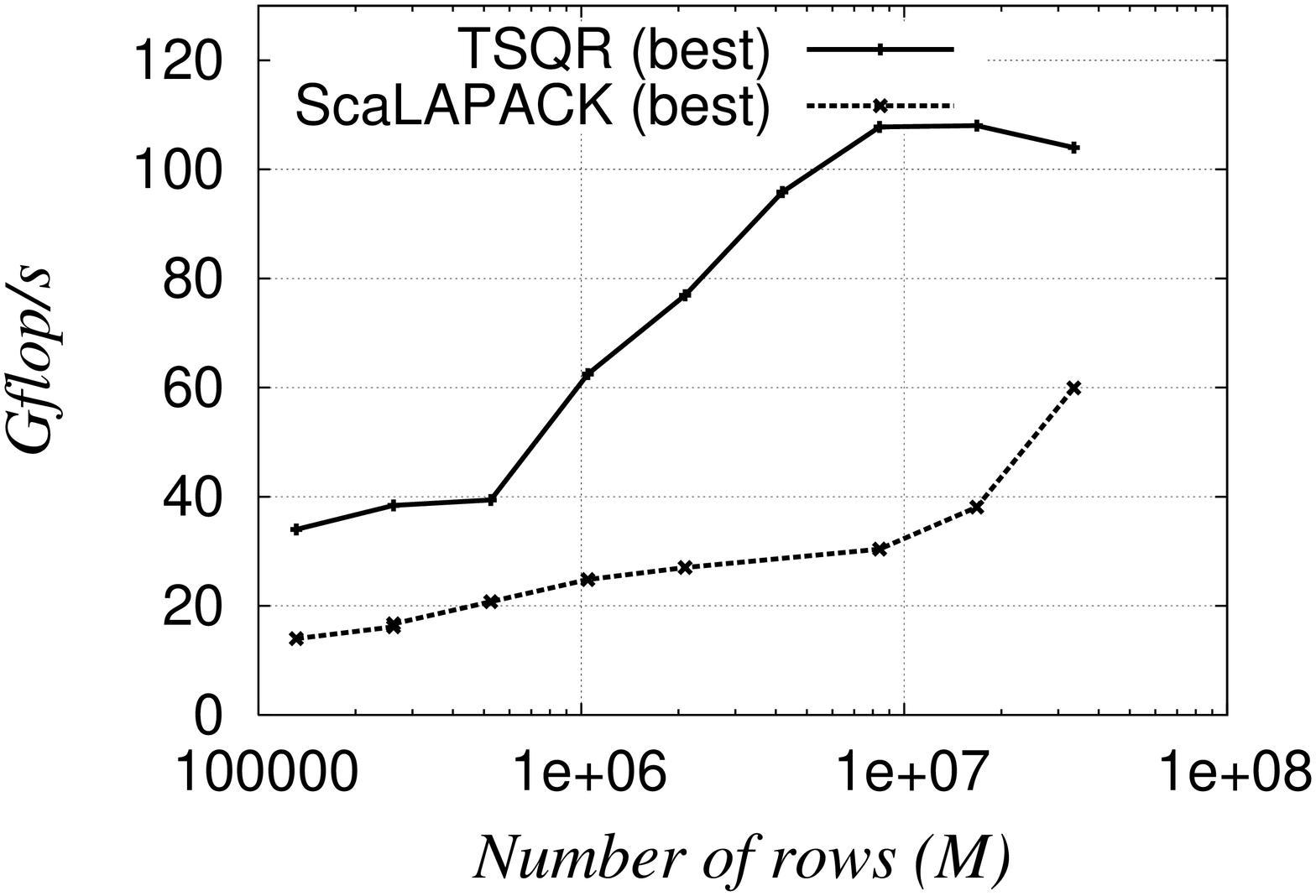}
    \label{fig:TSQR_QR_nb2}
  }\\
  \subfigure[N = 256]{
    \includegraphics[angle=270,width=\figurewidth]{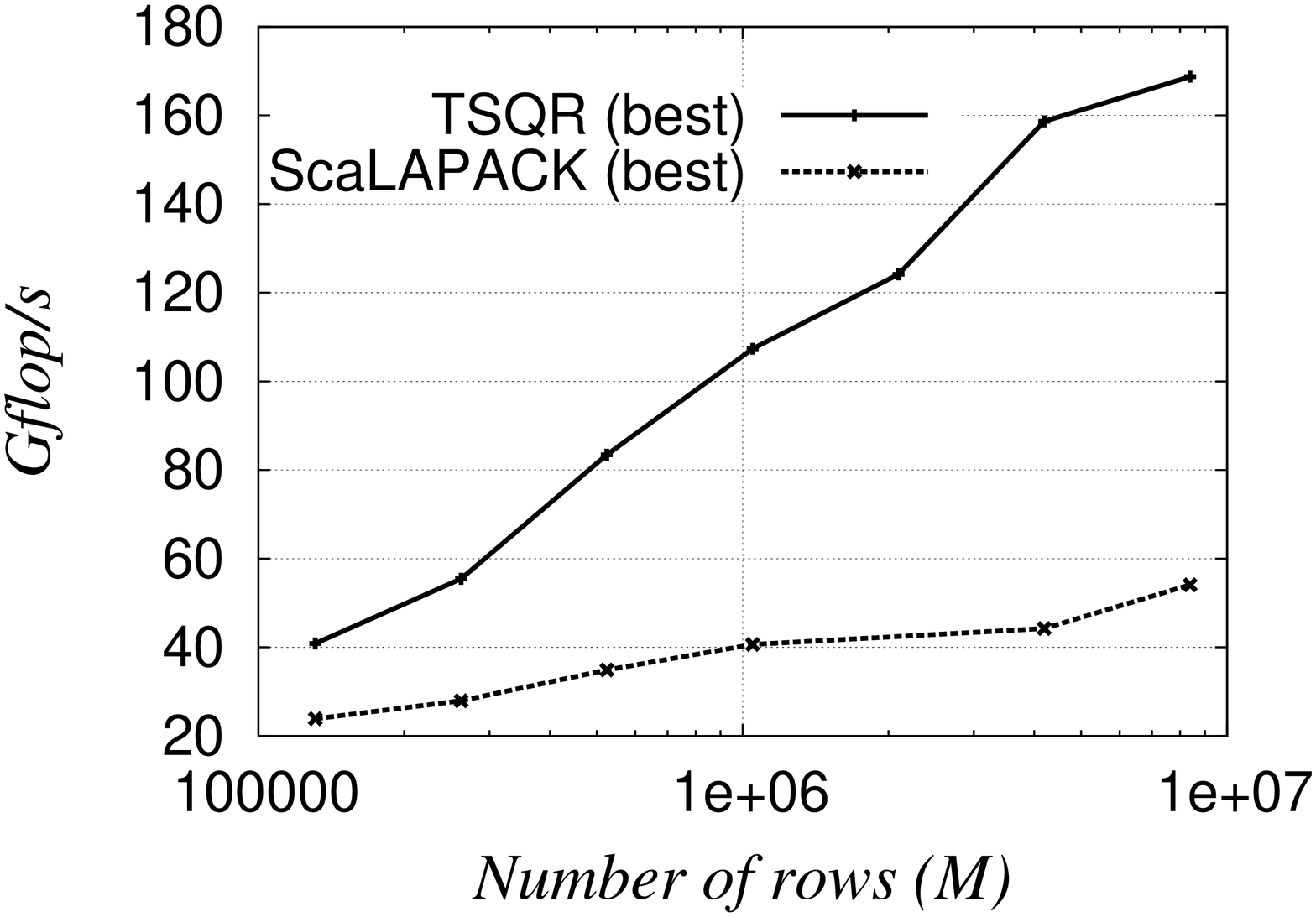}
    \label{fig:TSQR_QR_nb4}
  }
  \subfigure[N = 512]{
    \includegraphics[angle=270,width=\figurewidth]{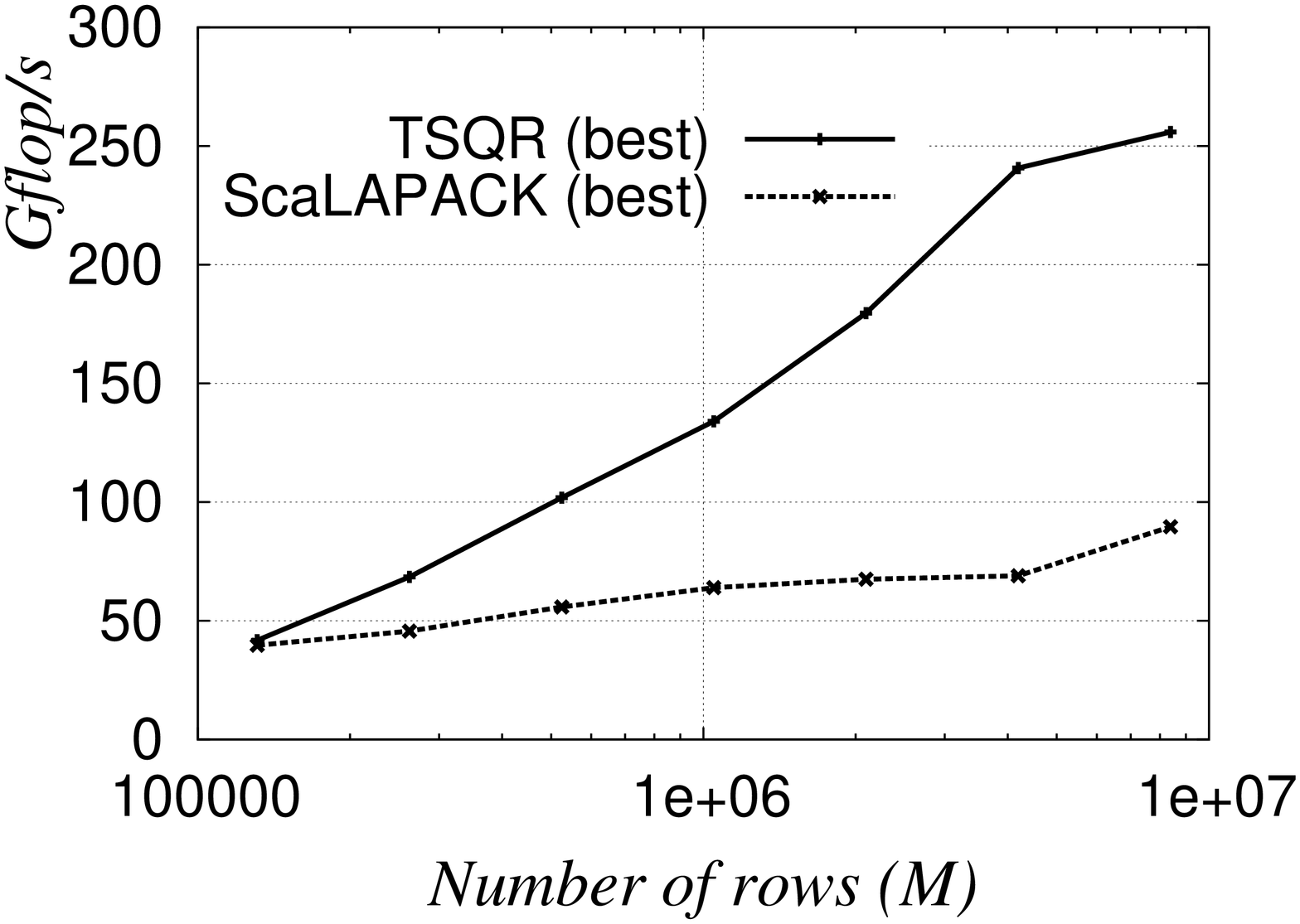}
    \label{fig:TSQR_QR_nb8}
  }
  \caption{TSQR vs ScaLAPACK. For each algorithm, the performance of
    the optimum configuration (one, two or four sites) is reported.}
  \label{fig:TSQR-vs-SCALA-2p1t}
\end{figure}

Figure~\ref{fig:TSQR-vs-SCALA-2p1t} compares TSQR performance (still
articulated with QCG-OMPI) against ScaLAPACK's.
We report the maximum performance out of executions on one, two or four
sites. For instance, the graph of TSQR in Figure~\ref{fig:TSQR_QR_nb1}
is thus the convex hull of the three graphs from
Figure~\ref{fig:TSQR_nb1}. In accordance with
Property~\ref{prop:tsqrbetter}, TSQR consistently achieves a higher
performance than ScaLAPACK. For matrices of limited height ($M =
131,072 $), TSQR is optimum when executed on one site
(Figure~\ref{fig:TSQR_nb1}). In this case, its superiority over
ScaLAPACK comes from better performance within a cluster
(Figure~\ref{fig:TSQR_1site_domaines_nb1}). For matrices with a larger
number of rows ($M=4,194,304$), the impact of the number of domains
per cluster is less sensitive
(Figure~\ref{fig:TSQR_1site_domaines_nb1} and Property~\ref{prop:m}).
On the other hand, the matrix is large enough to allow a speed up of
TSQR over the grid (Figure~\ref{fig:TSQR_nb1} and
Property~\ref{prop:m} (again)) but not of ScaLAPACK
(Figure~\ref{fig:QR_nb1} and Property~\ref{prop:tsqrbetter}), hence
the superiority of TSQR over ScaLAPACK for that type of matrix. For
very tall matrices ($M=33,554,432$), the impact of the number of
domains per cluster becomes negligible
(Figure~\ref{fig:TSQR_1site_domaines_nb1} and
Property~\ref{prop:m}). But (i) TSQR achieves a speed up of almost $4.0$
on four sites (Figure~\ref{fig:TSQR_nb1}) whereas (ii) ScaLAPACK 
does not achieve yet such an ideal speed up (Figure~\ref{fig:QR_nb1}).
Finally, on all the range of matrix shapes considered, and for
different reasons, we have seen that TSQR consistently achieves a significantly higher
performance than ScaLAPACK. For not so tall and not so skinny matrices
(left-most part of Figure~\ref{fig:TSQR_QR_nb8}), the gap between the
performance of TSQR and ScaLAPACK reduces
(Property~\ref{prop:tsqrbetter}).

One may have observed that the time spent in intra-node, then
intra-cluster and finally inter-cluster communications becomes
negligible while the dimensions of the matrices increase. For larger
matrices (which would not hold in the memory of our machines), we may
thus even expect that communications over the grid for ScaLAPACK would
become negligible and thus that TSQR and ScaLAPACK would eventually
achieve a similar (scalable) performance
(Property~\ref{prop:tsqrbetter}).

\ignore{
\subsection{1 proc 2 threads}
\label{sec:xp-vs}

\begin{figure}
  \centering
  \subfigure[TSQR]{
    \includegraphics[angle=270,width=\figurewidth]{TSQRscala2threads}
    \label{fig:TSQR2threads}
  }
  \subfigure[ScaLAPACK]{
    \includegraphics[angle=270,width=\figurewidth]{QRscala2threads}
    \label{fig:QR2threads}
  }
  \subfigure[Performance maximale]{
    \includegraphics[angle=270,width=\figurewidth]{TSQR_vs_QR}
    \label{fig:TSQRvsQR}
  }
  \caption[Passage à l'échelle des algorithmes TSQR et QR en utilisant deux
  threads par machine]{\label{fig:TSQRscala2threads}Passage à l'échelle de
    l'algorithme de factorisation QR à évitement de communications sur
    une matrice haute et fine sur une grille utilisant 1 à 4 grappes
    pour des matrices de un bloc en largeur et en faisant varier la
    hauteur, en utilisant deux threads par machine.}
\end{figure}

\subsection{Emulation of a grid with network controllers dedicated to
  processors}
\label{sec:xp-dedicated}

\begin{figure}
  \centering
  \subfigure[largeur = 1 bloc (64 doubles)]{
    \includegraphics[angle=270,width=\figurewidth]{TSQRscala}
    \label{fig:TSQRscala1}
  }
  \subfigure[largeur = 2 blocs (128 doubles)]{
    \includegraphics[angle=270,width=\figurewidth]{TSQRscala2}
    \label{fig:TSQRscala2}
  }\\
  \subfigure[largeur = 4 bloc (256 doubles)]{
    \includegraphics[angle=270,width=\figurewidth]{TSQRscala4}
    \label{fig:TSQRscala4}
  }
  \subfigure[largeur = 8 blocs (512 doubles)]{
    \includegraphics[angle=270,width=\figurewidth]{TSQRscala8}
    \label{fig:TSQRscala8}
  }
  \caption[Passage à l'échelle de l'algorithme TSQR sur une grille de 4
  grappes]{\label{fig:TSQRscala}Passage à l'échelle de l'algorithme de
    factorisation QR à évitement de communications sur une matrice
    haute et fine sur une grille utilisant 1 à 4 grappes pour
    différentes largeurs de la matrice et en faisant varier la
    hauteur.}
\end{figure}
}

\ignore{
\subsection*{}

This experimental study conducted on the Grid'5000 research platform
has shown that the performance of our approach increases linearly with
the number of geographical sites on large-scale problems (and is in
particular consistently higher than ScaLAPACK's).
}

\section{Conclusion and perspectives}
\label{sec:conclusion}

This paper has revisited the performance behavior of common dense
linear algebra operations in a grid computing environment. Contrary to
past studies, we have shown that they can achieve a performance speed
up by using multiple geographical sites of a computational grid.  To do
so, we have articulated a recently proposed algorithm (CAQR) with a
topology-aware middleware (QCG-OMPI) in order to confine intensive
communications (ScaLAPACK calls) within the different geographical
sites. Our experimental study, conducted on the experimental Grid'5000 platform,
focused on a particular operation, the QR factorization of TS
matrices. We showed that its performance increases linearly with the
number of geographical sites on large-scale problems (and is in
particular consistently higher than ScaLAPACK's).

We have proved theoretically through our models and experimentally
that TSQR is a scalable algorithm on the grid. TSQR is an important
algorithm in itself since, given a set of vectors, TSQR is a stable
way to generate an orthogonal basis for it. TSQR will come handy as an
orthogonalization scheme for sparse iterative methods (eigensolvers or
linear solves). TSQR is also the panel factorization of CAQR. A
natural question is whether CAQR scales as well on the grid. From
models, there is no doubt that CAQR should scale. However we will need
to perform the experiment to confirm this claim.  We note that the
work and conclusion we have reached here for TSQR/CAQR can be
(trivially) extended to TSLU/CALU (\cite{CALU}) and Cholesky
factorization~\cite{CACH}.

Our approach is based on ScaLAPACK. However, recent algorithms that
better fit emerging architectures would have certainly improved the
performance obtained on each cluster and \emph{in fine} the global
performance. For instance, recursive factorizations have been shown to
achieve a higher performance on distributed memory
machines~\cite{CAQR}. Other codes benefit from multicore
architectures~\cite{plasma-perf}.

If, as discussed in the introduction, the barriers for computational grids
to compete against supercomputers are multiple, this study shows that
the performance of large-scale dense linear algebra applications can
scale with the number of geographical sites. We plan to extend this work
to the QR factorization of general matrices and then to other
one-sided factorizations (Cholesky, LU). Load balancing to take into
account heterogeneity of clusters is another direction to
investigate. The use of recursive algorithms to achieve higher
performance is to be studied too.

\section*{Thanks}
The authors thank Laura Grigori for her constructive suggestions.


\footnotesize

\bibliographystyle{plain}
\bibliography{ipdps}

\end{document}